\documentclass[a4paper,11pt]{article}
\pdfoutput=1 
\RequirePackage{snapshot}  
\usepackage{jheppub} 
\usepackage{bm,amsmath,amssymb,slashed,graphicx,%
            enumerate,alltt,xspace,multirow,xcolor,mathrsfs}
\usepackage{fancyvrb}
\usepackage{booktabs,tabularx}
\usepackage{graphicx}
\usepackage{subcaption}
\usepackage{xspace}
\usepackage{cancel}
\usepackage[utf8]{inputenc} 
\usepackage[compat=1.0.0]{tikz-feynman}
\usepackage{scalerel}
\usepackage{layouts}
\usepackage{adjustbox}
\usepackage{pdflscape}
\usepackage[normalem]{ulem}
\usepackage{wasysym}
\usepackage{framed}
\usepackage{amsthm}
\usepackage[most]{tcolorbox}
\newtcbtheorem{observation}{Observation}%
{colback=white, colframe=black, fonttitle=\bfseries}{obs}
\newcommand{\ttt}[1]{\nolinkurl{#1}}

\usepackage{algorithm}
\usepackage[noEnd=true]{algpseudocodex}

\newcommand{\ps}{\text{\sc ps}}

\newcommand\betaps{\beta_\text{\textsc{ps}}}
\newcommand\betaobs{\beta_\text{obs}}

\usepackage[capitalise]{cleveref}

\makeatletter
\g@addto@macro\bfseries{\boldmath}
\makeatother

\definecolor{labelkey}{rgb}{0,0.5,0.0}
\definecolor{royalpurple}{rgb}{0.47, 0.32, 0.66}

\usepackage{listings}
\definecolor{darkred}{rgb}{0.6,0.0,0}
\definecolor{darkgreen}{rgb}{0,0.4,0}

\definecolor{grey}{rgb}{0.5,0.5,0.5}
\definecolor{rust}{rgb}{0.9,0.4,0.0}
\definecolor{lightblue}{rgb}{0.0,0.5,1.0}
\definecolor{ForestGreen}{rgb}{0.13,0.55,0.13}

\allowdisplaybreaks 

\definecolor{semiblue}{rgb}{0.3,0.3,0.8}
\newcommand{\logbook}[2]{}

\newcommand{\pgzerosdf}{PG$_{\beta=0}^\text{sdf}$\xspace}
\newcommand{\plhalf}{PL$_{\beta=1/2}$\xspace}

\newcommand{\order}[1]{\mathcal{O}\left(#1\right)}
\newcommand{\as}{\alpha_s}

\newcommand{\cH}{\mathcal{H}}
\newcommand{\cJ}{\mathcal{J}}
\newcommand{\cM}{\mathcal{M}}
\newcommand{\cP}{\mathcal{P}}
\newcommand{\cS}{\mathcal{S}}
\newcommand{\cW}{\mathcal{W}}

\newcommand{\nc}{N_\text{\textsc{c}}}

\newcommand{\rG}{r_{{\scriptscriptstyle \mathrm{G}}}}

\newcommand{\kT}{k_t}
\newcommand{\lnv}{\text{ln}v}

\newcommand\sss{\mathchoice%
{\displaystyle}%
{\scriptstyle}%
{\scriptscriptstyle}%
{\scriptscriptstyle}%
}

\newcommand{\rl}{r_{\rm \sss L}}
\newcommand{\rg}{r_{\rm \sss G}}
\newcommand{\spac}{{\hspace{0.5mm}}}

\PassOptionsToPackage{hypertexnames=false}{jheppub}

\tikzstyle{block} = [rectangle, minimum width=1.0cm, minimum height=0.75cm, thin, draw=black]
\tikzstyle{blob} = [circle, minimum width=0.5cm, thin, draw=black]
\tikzset{blackarrow/.style={-stealth, semithick, draw=black}}
\tikzset{connection/.style={inner sep=0,outer sep=0}}

\newcolumntype{C}{>{\centering\arraybackslash}X}

\title{The double-real part of NNLO matching with logarithmically accurate showers for $e^+e^-$ collisions}

\newcommand{\OXaff}{Rudolf Peierls Centre for Theoretical Physics,
  Department of Physics, Parks Road,
  University of Oxford, Oxford OX1 3PU, UK}
\newcommand{\ASCaff}{All Souls College, Oxford OX1 4AL, UK}
\newcommand{\NIKHEFaff}{Nikhef, Theory Group, Science Park 105, 1098 XG, Amsterdam, The Netherlands}
\newcommand{\CNRSaff}{Université Paris-Saclay, CNRS, CEA, Institut de physique théorique, 91191, Gif-sur-Yvette, France}
\newcommand{\UCLaff}{Department of Physics and Astronomy, University College London, London, WC1E 6BT, UK}

\author[a]{Melissa van Beekveld,}%
\author[b]{Keith Hamilton,}%
\author[c,d]{Gavin Salam,}%
\author[c]{Nicolas Schalch,}%
\author[e]{Gregory Soyez,}%
\author[c]{Silvia Zanoli}%

\affiliation[a]{\NIKHEFaff}
\affiliation[b]{\UCLaff}
\affiliation[c]{\OXaff}
\affiliation[d]{\ASCaff}
\affiliation[e]{\CNRSaff}

\preprint{{{\small Nikhef 2026-012, OUTP-26-11P}}}

\date{Received: date / Accepted: \today}

\abstract{
  We discuss key principles that need to be taken into account for
  higher-order matching to retain the logarithmic accuracy of parton
  showers, specifically in a context where the matching takes full
  responsibility for the first $n$ emissions.
  Failure to respect these principles breaks exponentiation and
  generates spurious leading-colour super-leading logarithms.
  Focusing on $e^+e^-$ collisions, and the PanScales showers, we construct
  a matching procedure that incorporates the leading-colour
  double-real, $\mathcal{O}(\as^2)$, contributions.
  It relies on new expressions for the
  inversion of these showers for any number of emissions,
  and extends techniques developed recently for the higher-order
  infrared limits of showers.
  Together, these ingredients allow us to use our showers as full
  phase-space generators.
  Our work establishes an important stepping stone towards
  logarithmically accurate NNLO matching.
}

\keywords{QCD, Parton Shower, Matching, Resummation, LEP
  \\[4em]
  \textit{For the purpose of Open Access, the authors have applied a CC BY
  public copyright licence to any Author Accepted Manuscript (AAM)
  version arising from this submission.}
}

\begin{document}
\maketitle

\section{Introduction}
\label{sec:intro}

A central goal of collider phenomenology is to understand fundamental interactions through precise predictions for experimentally measured observables. 
Fixed-order calculations provide a systematic expansion of the hard-scattering cross section in the strong coupling, but cannot capture the cumulative effect of arbitrarily many unresolved emissions. 
In soft and collinear regions, these emissions generate logarithmically enhanced corrections that must be resummed to all orders. 
In event simulations, this resummation is performed by parton showers, which evolve the hard process through successive branchings before being interfaced to a hadronisation model.
Reliable predictions across the full radiation phase space require fixed-order calculations and resummation to be combined through matching. 
The matching should reproduce the fixed-order description of hard, resolved radiation while preserving the all-order logarithmic structure in soft and collinear regions.

Fixed-order matching is well established at NLO through methods such
as MC@NLO~\cite{Frixione:2002ik} and
POWHEG~\cite{Nason:2004rx,Frixione:2007vw}, and has been generalised to
NNLO in extensively developed frameworks like \textsc{Geneva}~\cite{Alioli:2012fc,Alioli:2013hqa} and MiNNLO$_{\text{PS}}$~\cite{Hamilton:2012rf,Hamilton:2013fea,Monni:2019whf,Monni:2020nks}. 
Multi-jet merging combines matrix elements for several resolved multiplicities with a parton shower, as in CKKW~\cite{Catani:2001cc}, CKKW-L~\cite{Lonnblad:2001iq} and MLM~\cite{Mangano:2001xp}, with extensions to NLO such as \textsc{MePs@NLO}~\cite{Gehrmann:2012yg,Hoeche:2012yf}.
An alternative to separating multiplicities through a merging scale is
to correct successive shower emissions directly and multiplicatively,
as developed in Vincia~\cite{Giele:2007di,Giele:2011cb}, which has
also been extended to NNLO~\cite{Campbell:2021svd,El-Menoufi:2024sys}. 

With the development of showers with demonstrated logarithmic
accuracy, Refs.~\cite{Dasgupta:2020fwr,Hamilton:2020rcu,
  Karlberg:2021kwr,Hamilton:2021dyz,vanBeekveld:2022zhl,vanBeekveld:2022ukn,FerrarioRavasio:2023kyg,vanBeekveld:2023chs,vanBeekveld:2026wnj,vanBeekveld:2023ivn}
from our PanScales collaboration, and similar developments by other
groups, e.g.\ \cite{Nagy:2020rmk,Forshaw:2020wrq,Herren:2022jej,Hoche:2024dee,Preuss:2024vyu,Helliwell:2025krf},
it is essential that this accuracy not be spoiled by matching.
In the context of matching one real emission,
Ref.~\cite{Hamilton:2023dwb} showed that matching with a
hard-emission-generator (HEG) can compromise the accuracy of
event-shape distributions at next-to-next-to-double-logarithmic (NNDL)
accuracy, break exponentiation, and introduce spurious super-leading
logarithms (SLLs) in single-logarithmic
observables.\footnote{%
  Subleading-colour SLLs~\cite{Forshaw:2006fk} associated with Glauber 
  exchanges between initial states have been investigated extensively; 
  see, e.g., Refs.~\cite{Becher:2021zkk,Becher:2023mtx,Schalch:2026xxk}. 
  They are a manifestation of the broader phenomenon of coherence violation, 
  whose associated logarithms have been studied for a range of hadron-collider 
  observables~\cite{Banfi:2025mra,Dasgupta:2025cgl,Becher:2026kbr,Barcaro:2026dsd,Chen:2026dnj}.
  These developments are of crucial importance for the field, but it
  is just as important that we learn how to avoid introducing spurious
  leading-colour SLLs in the most widely used tools in collider
  phenomenology.  }
As we shall articulate here, the key to avoiding such problems more
generally is to ensure perfect alignment 
between the behaviours of the HEG and of the shower everywhere in the
infrared.

We explore the use of these principles --- and certain key limitations
--- in the context of matching the first two real emissions of a
shower.
The approach that we take is to use the kinematic maps and ordering
variables of existing logarithmically accurate PanScales (PanGlobal,
PanLocal) showers as the starting point for matching.
Our approach bears significant similarities with the Vincia
approach, in that we let the shower generate an emission, and
then construct an acceptance probability so as to reproduce the exact
real matrix element.
This is a form of nested multiplicative matching.
One substantial advantage of multiplicative matching for real
radiation is that it avoids negative-weight events.
Two essential new technical developments that we present here to make
that possible are expressions to invert one or more steps of the
shower, and the calculation of Jacobians to relate shower variables to
Lorentz-invariant phase space.\footnote{The Jacobians for the
  PanGlobal shower were recently calculated, independently, in
  Ref.~\cite{Holguin:2026trd}.
  We confirm that the expressions of that reference coincide with
  ours.  }

Multiplicative matching can have issues around coverage of phase space
and boundedness of reweighting factors.
The PanGlobal and PanLocal shower maps do cover phase space, but not
after imposing a shower ordering condition.
Furthermore, even in regions where phase space is covered, the
reweighting factors can grow large.
The second of these problems was already encountered in the context of
double-soft matching~\cite{FerrarioRavasio:2023kyg,2026PLDoubleSoft}
and here we explore the extension of some of the corresponding
technical advances to matching in full double-emission phase space.
One of the advances involves a specific, logarithmically safe,
prescription for adding disordered emissions~\cite{2026PLDoubleSoft},
which applies directly also in the full phase space.
The other advance involves simultaneously considering multiple
permutations of momenta and/or flavours (``swaps'' in the language of
Ref.~\cite{FerrarioRavasio:2023kyg}), which we here develop further,
notably for the PanLocal shower.

The use of identical techniques for hard matching and infrared
matching is a key step towards maintaining logarithmic accuracy and
avoiding SLLs.
One important caveat is that, currently, neither the PanScales showers nor
any other showers incorporate the correct triple-collinear splitting
structure, and thus they won't align with the  double-real matrix elements
in the corresponding limit.
As we shall argue below, until logarithmically accurate showers become
available with full double-soft \emph{and} triple-collinear structure
in the IR, there will inevitably remain sources of SLLs in matching
the hard double-emission structure.

Using the above set of techniques, we construct a nested
multiplicative matching of the PanScales showers to the first two real emissions in $e^+e^-$ collisions. 
The matching reproduces the exact leading-colour tree-level matrix elements for four-parton final states, corresponding to the complete leading-colour double-real contribution at $\mathcal{O}(\as^2)$, while preserving the shower’s double-soft evolution. 
The resulting shower is simultaneously NLO 2-jet, LO 3-jet and LO 4-jet
accurate for unoriented events.
This constitutes an essential step towards NNLO+NNLL matching, noting that the real-virtual and double-virtual contributions required for full NNLO accuracy are beyond the scope of this work.

The structure of the paper is as follows.
Section~\ref{sec:matching-log-accuracy} introduces the key principle
for matching to logarithmically accurate showers without introducing
spurious SLLs.
Section~\ref{sec:shower-inversion} presents the PanScales kinematic
maps, inversions and Jacobians that are needed for our matching procedure.
Section~\ref{sec:match-proc} presents the nested matching procedure,
including phase-space coverage, multiplicative correction factors and
the assignment of flavour and colour channels.
Section~\ref{sec:validation} validates the fixed-order expansion
against \texttt{Event2}~\cite{Catani:1996jh,Catani:1996vz},
examines the double-soft limit, carries out tests of logarithmic
accuracy, and discusses where matching-induced SLLs remain as a result
of the lack of triple-collinear terms in the shower.
We conclude in Section~\ref{sec:conclusions}.
Phenomenological results will be discussed in a forthcoming companion paper,
together with a discussion of shower uncertainties~\cite{PanPheno}.

\section{Matching and logarithmic accuracy}
\label{sec:matching-log-accuracy}

There are two broad classes of higher-order matching that are in use:
additive matching, like MC@NLO~\cite{Frixione:2002ik}\footnote{First steps towards 
  additive NNLO matching have recently been published
  in~\cite{Campbell:2026aqe}.}
and HEG-style
matching like POWHEG~\cite{Nason:2004rx}.
It has long been obvious that for additive matching to be
straightforward, the shower must exactly reproduce the singular limits
at the given order of matching.
Here we focus on HEG-style matching, where the importance of perfect
alignment between any matching procedure and the parton shower in
infrared limits has been discussed, in the context of logarithmic
accuracy, only for NLO matching~\cite{Hamilton:2023dwb}.
However, in broad terms, the same kind of issue is expected to arise
beyond NLO.
Specifically the following key principle needs to be respected:

\begin{observation}{key matching principle}{key-matching}
  In any scheme where the matching takes full responsibility for the
  first $n$ emissions, the soft and/or collinear limits for those $n$
  emissions must be engineered to {\it exactly} reproduce the way the shower
  would produce those $n$ emissions.
\end{observation}
\noindent If that principle is not satisfied, the combination of shower and matching is almost
  guaranteed to generate spurious leading-colour super-leading logarithms for
  some observables. 
  There is also a corollary:
\begin{observation}{corollary, constraints on parton showers}{corollary}
  In any scheme where the matching takes full responsibility for the
  first $n$ emissions, the shower must itself embody the complete
  singular limits for any $n$ soft and/or collinear emissions.
\end{observation}

\begin{figure}
  \centering
  \includegraphics[width=0.8\textwidth]{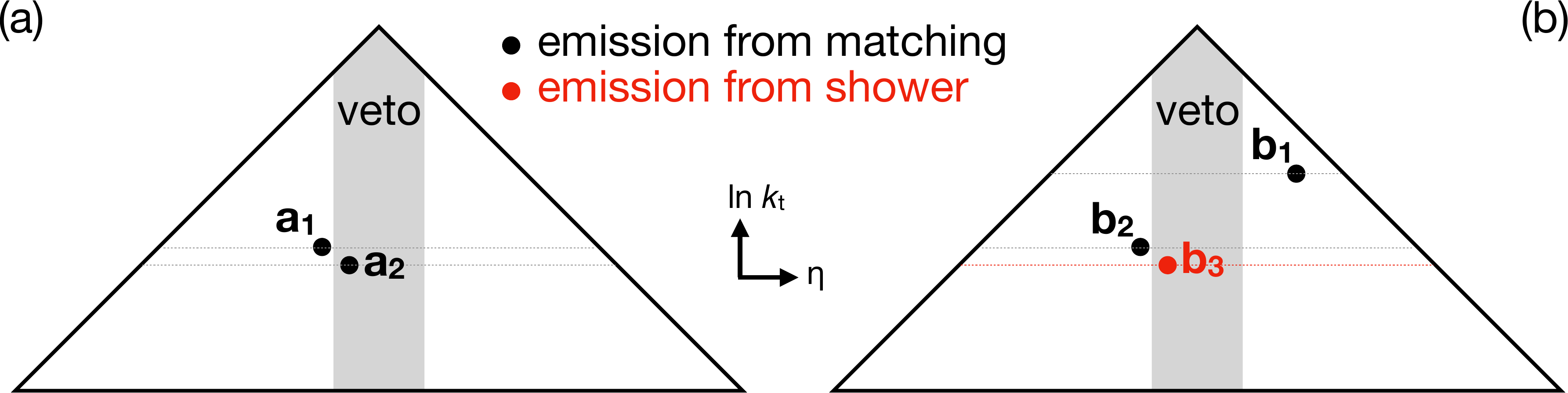}
  \caption{Illustration of the need for full alignment in the
    infrared between the matched generation of the first two emissions
    (left, $a_2$ given $a_1$) and the shower's generation of the corresponding
    two-emission phase space (right, $b_3$ given $b_2$).
    For an emission veto in the slice (shown in grey),
    any leading-power mismatch results in spurious super-leading logarithms, because
    the probability of triggering the mismatch is connected with the
    double-logarithmic probability of emitting $b_1$.
    See text for further details.
  }
  \label{fig:ds-avoid-SLLs}
\end{figure}

To help illustrate this, consider Fig.~\ref{fig:ds-avoid-SLLs}, where
the observable that we keep in mind is a veto on energy flow into a
rapidity slice (with the slice shown in grey).
On the left-hand side, we have two emissions generated by
matching with the four-parton matrix element: $a_1$ just outside the
vetoed slice region and $a_2$ just inside.
Let's suppose that their transverse momenta are small,
$k_t \sim e^{L}$, with large negative $L$.
Having the correct distribution of these two emissions is a key
component for next-to-single-logarithmic accuracy for a veto on
the energy-flow in a slice.
On the right-hand side, Fig.~\ref{fig:ds-avoid-SLLs}b, we again have
two matched emissions, $b_1$ and $b_2$: $b_1$ is now harder than $a_1$, but soft and
collinear, and $b_2$ plays the role of $a_1$ in
Fig.~\ref{fig:ds-avoid-SLLs}a.
The pair that straddles the boundary of the vetoed region involves
one matched emission, $b_2$, and one double-soft shower emission, $b_3$.
Because of factorisation, the true QCD probability distributions of
the $a_1a_2$ pair on the left and the $b_2 b_3$ pair on the right are
identical: there is no sensitivity to soft-collinear emissions
elsewhere in the event, as long as such emissions are far away in the
Lund plane.

The issue that concerns us is what happens if the combination of
matching and parton shower results in a mismatch at some relative
order $\as^p$ between the treatment of the pair straddling the
boundary in the left- and right-hand diagrams.
Schematically, suppose that in both diagrams the matched shower (or,
equivalently, a HEG) has reached
the scale associated with the emission just outside the veto region.
Given that state, on the left, let us write the probability that the
matched shower or HEG now produces the $a_1,a_2$ pair as $P_\text{matched}(a_1,a_2)$.
Given the state on the right, the probability of
producing the corresponding $b_2,b_3$ pair is
$P_\text{shower}(b_2,b_3)$ (the matching has no impact on the
generation of $b_2$, because it is far in the Lund plane from $b_1$,
and it is the shower's conditional probability
that determines the generation of $b_3$).
Now suppose we have a mismatch between those two conditional
probabilities, $P_\text{matched}(a_1,a_2) = (1 + C  \as^p  )
P_\text{shower}(b_2,b_3)$, where 
$C$ is some number of order $1$ and the power $p$ depends on the
origin of the mismatch. 
The probabilities $P$ are to be understood as not including Sudakov
form factors.
Taking those into account, in a double-logarithmic approximation, we
have a probability
$P_\text{matched}(a_1,a_2) e^{-\frac{2C_F\as}{\pi} L^2 }$ for
generating $a_1$ and $a_2$;
and a probability
$P_\text{shower}(b_2,b_3) \left(1 - e^{-\frac{2C_F\as}{\pi} L^2
  }\right)$ for producing $b_2$ and $b_3$ inclusive over any number of
previous soft-collinear emissions.
Thus the overall probability for having the two soft large-angle emissions
straddling the slice boundary is
\begin{multline}
  \label{eq:sll-issue}
  P_\text{matched}(a_1,a_2) e^{-\frac{2C_F\as}{\pi} L^2 } +
  P_\text{shower}(b_2,b_3) \left(1 - 
    e^{-\frac{2C_F\as}{\pi} L^2 }\right)\\
  = P_\text{shower}(b_2,b_3)
  \left(1 +
    C \as^p 
    e^{-\frac{2C_F\as}{\pi} L^2 }
  \right).
\end{multline}
The key point to note is that no matter how large $p$ is, the
$e^{-\frac{2C_F\as}{\pi} L^2 }$ factor inevitably results in
terms $\as^n L^m$ with $m > n$.
Such terms would be super-leading relative to the known
single-logarithmic resummation for energy in a slice, and therefore
spurious.
Furthermore, unlike the physical super-leading logarithms in processes
with initial-state hadrons~\cite{Forshaw:2006fk,Becher:2021zkk}, they
would appear at leading colour.
Their phenomenological impact will depend on the values of $p$ and $C$.
However, generically, the presence of double-logarithmic terms in an
observable that is normally just single-logarithmic is expected to
introduce non-trivial spurious structure in the region
$\as L^2 \sim 1$, as was seen for example in
Ref.~\cite{Hamilton:2023dwb}, notably Fig.~9 (bottom-right) for the
single-logarithmic Soft-Drop observable.

Note that for double-logarithmic observables like event shapes, the
analogue of the problem discussed above for the rapidity slice is that a
misalignment between matching and shower can break exponentiation.
This is discussed below in Section~\ref{sec:log-accuracy-tests}.

The task we set ourselves in the next sections is to explore
multiplicative matching in such a way as to respect the key principle of
box~\ref{obs:key-matching} above.
The matching developed in this work
achieves this alignment at leading power and leading colour in the
double-soft limit: $P_{\rm matched}/P_{\rm shower}\to1$, so the leading-power
mismatch coefficient $C$ in Eq.~\eqref{eq:sll-issue} vanishes, thereby
avoiding the generation of unphysical SLLs via the mechanism sketched in
Fig.~\ref{fig:ds-avoid-SLLs}.
There will be one important caveat to this work: the PanScales showers
that we are matching currently incorporate the double-soft limit, but
not the triple-collinear limit, which violates the corollary
(box~\ref{obs:corollary}) 
above.
From that point of view, the work here should be taken not as the
final word on 4-parton matching, but rather as a powerful proof of
principle, illustrating some of the key considerations that arise.
It also prepares the ground for a complete solution once there are
showers that also incorporate the triple-collinear limit.
Despite this caveat, what we achieve here will already be a
considerable advance in terms of matching and logarithmic accuracy, and
its understanding relative to the literature.

\section{PanScales shower maps, inversions and Jacobians}
\label{sec:shower-inversion}

An essential ingredient in multiplicative matching is to be able to
reconstruct all branching paths through which the shower could have
reached a given phase-space point.
This is possible only if the shower maps are fully invertible: for any
choice of emitter, spectator and radiation, the post-branching momenta
determine an underlying pre-branching configuration, and a set of
corresponding branching variables. Ideally, this reconstruction is
analytic and unique throughout the physical domain. This procedure
also requires knowledge of the phase-space Jacobian associated with
branchings.

We use the PanLocal and PanGlobal shower maps of
Refs.~\cite{Dasgupta:2020fwr,FerrarioRavasio:2023kyg},  
which we briefly review before presenting their inversions.
While the ability to invert 
these maps analytically is a key ingredient of our matching procedure,
the details are rather technical, and readers primarily interested in the 
matching methodology may skip this section and directly go to Section
\ref{sec:match-proc}.
Simply knowing that one can express all pre-branching quantities and
the phase-space Jacobian in terms of post-branching kinematic
invariants is all that will be needed to understand our matching
procedure.

\subsection{PanGlobal shower}
\label{sec:shower-inversion-pglr}

For a given candidate branching, we use the indices $i$, $j$ and $k$ to label 
the emitter, spectator and radiated parton, respectively. Throughout this 
section, tilded momenta denote the pre-branching configuration, barred momenta 
the intermediate configuration obtained from the local $2\to3$ map,
and undecorated symbols represent 
momenta in the physical post-branching configuration. Thus, for example, $\tilde p_i$ 
and $\tilde p_j$ denote the pre-branching emitter and spectator, while $\bar p_i$, 
$\bar p_j$, $\bar p_k$ and $p_i$, $p_j$, $p_k$ denote the corresponding intermediate 
and post-branching momenta. For the latter, we define
\begin{align}\label{eq:PG-pd-pm-dsq}
p_d &\equiv p_i+p_j+p_k,
&
p_m &\equiv Q-p_d,
&
d^2 &\equiv p_d^2 \spac ,
\end{align}
where $Q$ is the total event momentum. Here, $p_m$ is the total momentum of 
all particles not directly involved in the branching. We furthermore introduce 
the notation
\begin{equation}\label{eq:sxy}
s_{xy}\equiv 2p_x\cdot p_y,
\qquad
s_{x}\equiv 2p_x\cdot Q,
\end{equation}
with analogous definitions for pre-branching and intermediate
momenta.

\subsubsection{Forward map}\label{sec:pg-forward-map}
The PanGlobal map is most conveniently described in two steps. Starting from a 
dipole composed of $\{\tilde p_i , \tilde p_j\}$ that is radiating an emission $p_k$,
the dipole-local \(2\to3\) map constructs the intermediate momenta
\begin{subequations}
  \begin{align}
    \bar p_k &=
               a_k\tilde p_i+b_k\tilde p_j-k_\perp\,,
               \label{eq:PG-local-pk}
    \\
    \bar p_i &=
               (1-a_k)\spac \tilde p_i\,,
               \label{eq:PG-local-pi}
    \\
    \bar p_j &=
               (1-b_k)\spac \tilde p_j\, ,
               \label{eq:PG-local-pj}
  \end{align}\label{eq:PG-local}
\end{subequations}
where the transverse vector satisfies
\begin{equation}
  \label{eq:PG-kperp-def}
  k_\perp\cdot\tilde p_i
  =
  k_\perp\cdot\tilde p_j
  =
  0\,,
  \qquad
  -k_\perp^2=a_kb_k\,s_{\tilde \imath \tilde \jmath} \equiv k_t^2\,,
\end{equation}
so that all three intermediate momenta are massless.
The map in Eq.~(\ref{eq:PG-local}) breaks momentum conservation, which
we then restore through a ``local'' rescaling ($\rl$) of just the
dipole momenta, an optional global rescaling of the whole event
($\rg$), and a boost of the whole event.
We return to these below.

A common feature among PanScales
showers is the ordering variable $v$, which, regardless of the recoil assignment, is defined
as
\begin{equation}
  \label{eq:evolution-variable}
  \kT = \rho v e^{\betaps|\bar \eta|}\,,
  \qquad
  \rho = \left(\frac{s_{\tilde{\imath} }
      s_{\tilde{\jmath} }}{Q^2 s_{\tilde{\imath}\tilde{\jmath}}}\right)^{\frac{\betaps}{2}}\, ,
\end{equation}
with a continuous parameter $\betaps$, ranging from $0\leq \betaps < 1$. The choice $\betaps=0$
corresponds to transverse momentum ordering.
Note that sometimes, especially in figures, we will
use the notation $\beta$ instead of $\betaps$.
The light-cone components of the emission $p_k$
are
\begin{equation}
  a_{k}\equiv\sqrt{\frac{s_{\tilde{\jmath}}}{s_{\tilde{\imath}\tilde{\jmath}}s_{\tilde{\imath}}}}\,
  \kT{e}^{+\bar \eta}\,,
  \qquad
  b_{k}\equiv\sqrt{\frac{s_{\tilde{\imath}}}{s_{\tilde{\imath}\tilde{\jmath}}s_{\tilde{\jmath}}}}\,
  \kT{e}^{-\bar \eta}\,,
  \label{eq:ak-bk-Sudakov-coeffs}
\end{equation}
where the rapidity $\bar \eta$ is defined in the event centre-of-mass frame, 
and can be related to the rapidity $\eta$ in the dipole frame through
\begin{align}
  \bar \eta = \eta + \frac{1}{2}\ln \frac{s_{\tilde \imath}}{s_{\tilde \jmath}} \spac .
\end{align}
More precisely, $\eta = 0$ bisects the dipole in its rest frame, while $\bar \eta = 0$ 
does so in the event frame.
Defining
\begin{equation}
  \bar p_d\equiv\bar p_i+\bar p_j+\bar p_k\,,
  \qquad
  \tilde p_m\equiv
  Q-\tilde p_i-\tilde p_j
  =
  \sum_{l\notin\{i,j,k\}}\tilde p_l\,,
  \label{eq:pd-and-pm-defns}
\end{equation}
the total momentum prior to the event-wide Lorentz transformation
is
\begin{equation}
  \bar Q\equiv \rl\spac \bar p_d+\tilde p_m\,.
\end{equation}
A possible additional global rescaling is defined by 
\begin{equation}
  \rG
  =
  \sqrt{\frac{Q^2}{\bar Q^2}}\,,
  \qquad
  Q'=\rG\bar Q\,.
\end{equation}
The second step is to boost the physical
post-branching momenta
\begin{subequations}
\begin{align}
  p_a &=
  \mathbb B\,
  \rG\rl\spac \bar p_a\,,
  && a\in\{i,j,k\}\,,
  \label{eq:PG-variants-pijk}
  \\
  p_l &=
  \mathbb B\,
  \rg\spac \tilde p_l\,,
  && l\notin\{i,j,k\}\,.
  \label{eq:PG-variants-pl}
\end{align}
\end{subequations}
Here, $\mathbb B$ is a Lorentz transformation that maps
$Q'$ onto $Q$. It is written as
\begin{equation}
  \mathbb B^\mu{}_\nu
  =
  g^\mu{}_\nu
  +\frac{2Q^\mu Q'_\nu}{Q^2}
  -\frac{2(Q+Q')^\mu(Q+Q')_\nu}{(Q+Q')^2}\,,
  \label{eq:PG-variants-Boost}
\end{equation}
where $Q'^2=Q^2$, and therefore
$\mathbb B^\mu{}_\nu \spac Q'^\nu=Q^\mu$.

In this work, we use the (default) \ttt{LocalRescaling} variant of the
PanGlobal map. In this variant, $\rl$ is chosen such that
$\bar Q^2=Q^2$, so that no subsequent global rescaling is required,
\begin{equation}
  \rG=1\,.
\end{equation}
Therefore, we have
\begin{equation}
  \rl
  =
  \frac{
    -\tilde p_m\cdot\bar p_d
    +
    \sqrt{
      (\tilde p_m\cdot\bar p_d)^2
      +
      \bar p_d^2\bigl(Q^2-\tilde p_m^2\bigr)
    }
  }{
    \bar p_d^2
  }\,.
  \label{eq:rL-PG-forward}
\end{equation}
The positive root is selected so that the rescaling is continuously
connected to \(r_{\mathrm L}=1\) in the unresolved limit.\\
\subsubsection{Inversion}\label{sec:pg-inversion}
Having completed the construction of the forward shower map, we 
now turn to its inversion. Starting from the post-branching momenta 
and a given assignment of the emitter, spectator and radiation, we 
determine the corresponding pre-branching configuration. 
As a first step, the local rescaling factor can be written as
\begin{equation}
\rl =
\frac{-\beta_{{\scriptscriptstyle \mathrm{L}}}+\sqrt{\beta_{{\scriptscriptstyle \mathrm{L}}}^{2}+4\alpha_{{\scriptscriptstyle \mathrm{L}}}\gamma_{{\scriptscriptstyle \mathrm{L}}}}}{2\alpha_{{\scriptscriptstyle \mathrm{L}}}}\,,
\label{eq:rL-PG}
\end{equation}
where
\begin{subequations}\label{eq:ee-PGLR-rL-gamma}
\begin{align}
\alpha_{{\scriptscriptstyle \mathrm{L}}}\,=\,& \frac{1}{2}\,s_{ij}\,\left(\,2s_{d}-s_{dd}\,\right)\,,\\
\beta_{{\scriptscriptstyle \mathrm{L}}}\,=\,&2s_{id}\,s_{jd}-s_{i}\,s_{jd}-s_{j}\,s_{id}\,,\\
\gamma_{{\scriptscriptstyle \mathrm{L}}}\,=\,& s_{id}\,s_{jd}\,.
\end{align}
\end{subequations}
The light-cone components $a_k$ and $b_k$ are given by
\begin{equation}
a_{k}\,=\, \frac{s_{jk}}{s_{ij}+s_{jk}}\,,\qquad
b_{k}\,=\, \frac{s_{ik}}{s_{ij}+s_{ik}}\, .
\end{equation}
With these ingredients at hand, we can construct the pre-branching scalar products involving the dipole momenta $\tilde \imath$ and $\tilde \jmath$ as
\begin{subequations}\label{eq:ee-PGDR-tilde-sjQ}
\begin{align}
  s_{\tilde{\imath}\tilde{\jmath}}\,=\, & \frac{1}{
                                          \rl^{2}}\,\frac{1}{s_{ij}}\,(s_{ij}+s_{ik})\,(s_{ij}+s_{jk})\,,\\
  s_{\tilde{\imath}}\,=\, & \frac{1}{
                             \rl^{2}}\,\frac{1}{s_{ij}}\,(s_{ij}+s_{jk})\,(s_{ij}+s_{ik}+\rl\,s_{im})\,,\\
  s_{\tilde{\jmath}}\,=\, & \frac{1}{
                             \rl^{2}}\,\frac{1}{s_{ij}}\,(s_{ij}+s_{ik})\,(s_{ij}+s_{jk}+\rl\,s_{jm})\,.
\end{align}
\end{subequations}
Moreover, considering an arbitrary momentum
$\tilde l$ external to the dipole,  the complete set of
pre-branching scalar products is then given by
\begin{subequations}\label{eq:ee-PGDR-tilde-sllprime}
\begin{align}
  s_{\tilde{\imath}\tilde{l}}\,=\, & \frac{1}{
                                     \rl}\,\left(\frac{1}{1-a_{k}}\,s_{il}\right),\\
  s_{\tilde{\jmath}\tilde{l}}\,=\, & \frac{1}{
                                     \rl}\,\left(\frac{1}{1-b_{k}}\,s_{jl}\right),\\
  s_{k_{\perp}\tilde{l}}\,=\, & \frac{1}{
                                \rl}\,\left(\frac{a_{k}}{1-a_{k}}\,s_{il}+\frac{b_{k}}{1-b_{k}}\,s_{jl}-s_{kl}\right),\\
  s_{\tilde{l}}\,=\, & \frac{1}{
                        \rl}\,\left[\left(\frac{1}{1-a_{k}}-\rl\right)s_{il}+\left(\frac{1}{1-b_{k}}-\rl\right)s_{jl}+\rl\left(s_{l}-s_{kl}\right)\right],\\
  s_{\tilde{l}\tilde{l}^{\prime}}\,=\, & 
                                         s_{ll^{\prime}}\,.
\end{align}
\end{subequations}
In practice, the matching procedure described in
Section~\ref{sec:match-proc} requires only the above pre-branching scalar
products. For completeness, however,
we also provide the reconstructed momenta, which may be useful in
future applications requiring the full pre-branching event. We begin
by defining the following auxiliary scalar quantities:
\begin{subequations}
\begin{align}
\xi_{i}\,& =\,\frac{2s_{\tilde{\imath}}+s_{\tilde{\imath}\tilde{\jmath}}\,(\rl-1)}{(Q+\bar{Q})^{2}}\,,\\[2mm]
\xi_{j}\,& =\,\frac{2s_{\tilde{\jmath}}+s_{\tilde{\imath}\tilde{\jmath}}\,(\rl-1)}{(Q+\bar{Q})^{2}}\,, \label{eq:ee-PGLR-xi-i-xi-j} \\[2mm]
\xi_{l}\,& =\,\frac{2s_{\tilde{l}}+\rl(s_{\tilde{\imath}\tilde{l}}+s_{\tilde{\jmath}\tilde{l}}-s_{k_{\perp}\tilde{l}})-(s_{\tilde{\imath}\tilde{l}}+s_{\tilde{\jmath}\tilde{l}})}{(Q+\bar{Q})^{2}}\,,\label{eq:ee-PGLR-xi-l} \\[2mm]
(Q+\bar{Q})^{2}\,& =\,4Q^{2}\,+\,s_{\tilde{\imath}\tilde{\jmath}}\,\left[\,\rl^{2}\,a_{k}\,b_{k}-(\rl-1)^{2}\,\right]\,,\label{eq:PGLR-Q-plus-Qbar-sqd}
\end{align}
\end{subequations}
and the auxiliary vector
\begin{equation}
\mathbb{B}\,Q\,=\,\mathbb{B}\,(\tilde{p}_{i}+\tilde{p}_{j}+\tilde{p}_{m})\,=\,\frac{1}{\rl(1-a_{k})}\,p_{i}+\frac{1}{\rl(1-b_{k})}\,p_{j}+p_{m}\,.\label{eq:ee-PGLR-BQ}
\end{equation}
The pre-branching dipole momenta and $k_{\perp}$ can then be assembled
as 
\begin{subequations}
\begin{align}
\tilde{p}_{i}\,=\, & \frac{1}{\rl(1-a_{k})}\,p_{i}-\xi_{i}\,\mathbb{B}\,Q+\left(\frac{s_{\tilde{\imath}}}{Q^{2}}-\xi_{i}\right)Q\,,\label{eq:ee-PGLR-tilde-pi}\\
\tilde{p}_{j}\,=\, & \frac{1}{\rl(1-b_{k})}\,p_{j}-\xi_{j}\,\mathbb{B}\,Q+\left(\frac{s_{\tilde{\jmath}}}{Q^{2}}-\xi_{j}\right)Q\,,\label{eq:ee-PGLR-tilde-pj}\\
k_{\perp}\,=\, & \frac{\rl-1}{\rl}\,(\tilde{p}_{i}+\tilde{p}_{j})+\frac{1}{\rl}\mathbb{B}\,Q-\frac{1}{\rl}\left[\frac{(Q+\bar{Q})^{2}}{Q^{2}}-3\right]Q\, , \label{eq:ee-PGLR-kperp}
\end{align}
\end{subequations}
as well as all remaining pre-branching vectors
\begin{align}
\tilde{p}_{l}\,=\, & p_{l}-\xi_{l}\,\mathbb{B}\,Q+\left(\frac{s_{\tilde{l}}}{Q^{2}}-\xi_{l}\right)Q\,.\label{eq:ee-PGLR-tilde-pl-v2}
\end{align}
Finally, we can choose to compute $\bar{Q}$ from
\begin{equation}
\bar{Q}\,=\,Q+\left(\rl-1\right)\left(\tilde{p}_{i}+\tilde{p}_{j}\right)\,-\,\rl k_{\perp}\,.\label{eq:ee-PGLR-Qbar}
\end{equation}
\subsubsection{Jacobian}
\label{sec-pg-jac}
The role of the inverse map is not limited to reconstructing the
underlying momenta. The same variables that define the emission also
provide a parametrisation of the Lorentz-invariant phase-space
measure $\Phi$. For a single emission this can be written as
\begin{align}
  d\Phi_{n+1}(Q;\{p\}_{n+1})
  =
  d\widetilde{\Phi}_{n}(Q;\{\tilde p\}_{n})\,
  d\Phi_{\rm rad}, \label{eq:phsp-fac}
\end{align}
where the underlying momenta $\{\tilde p\}_{n}$ are obtained from the
inverse map. In terms of the shower variables, we write the radiation
phase space
\begin{align}
  d\Phi_{\rm rad}
  =
  \frac{1}{16\pi^2}\,
  da_k\,db_k\,\frac{d\phi_k}{2\pi}\times
  \mathcal{J} = \frac{1}{16\pi^2}\,
  d\ln v_k\,d\bar\eta_k\,\frac{d\phi_k}{2\pi}
  \times 2a_k b_k\spac \mathcal{J} .
  \label{eq:shower-rad-phsp}
\end{align}
The phase-space factorisation \eqref{eq:phsp-fac} holds for any
particle multiplicity $n$ and can therefore be iterated for an
arbitrary number of emissions.
For each branching, the corresponding Jacobian, $\cJ$, is fully determined by
the pre-branching scalar products and branching variables returned by
the inverse map. For the PanGlobal \ttt{LocalRescaling} map the Jacobian is
\begin{align}\label{eq:phsp-jac-pg-lr}
  \mathcal{J}\,=\,\rl^{6}\,s_{\tilde{\imath}\tilde{\jmath}}\,(1-a_{k})\,(1-b_{k})\,\frac{s_{\tilde{\imath}}+s_{\tilde{\jmath}}}{s_{\tilde{\imath}}+s_{\tilde{\jmath}}-s_{\tilde{\imath}\tilde{\jmath}}+s_{\tilde{\imath}\tilde{\jmath}}\,\rl^{2}\,(1-a_{k}b_{k})}.
\end{align}
The shower branching probability is naturally expressed as a density
in the shower variables
\begin{align}
  \frac{d\cP}{d\lnv_k \, d\bar\eta_k \, \frac{d\phi_k}{2\pi}} = 
  \sum_{\text{dipoles } \{\tilde{\imath},\tilde{\jmath}\}}
  \frac{\as(\kT) }{\pi} \times
  \left[f(\bar\eta_k) a_k P_{\tilde{\imath} \to ik}(a_k) +
    f(-\bar\eta_k) b_k P_{\tilde{\jmath} \to jk}(b_k)
    \right],\label{eq:dPh}
\end{align}
where $f(\bar \eta) = 1/(1+e^{-2\bar \eta})$ is a partitioning function, and 
$P_{\tilde{\imath} \to ik}(z)$ are the standard first-order splitting
functions~\cite{Gribov:1972ri,Altarelli:1977zs,Dokshitzer:1977sg}. 
For the PanGlobal shower, we use two different ways of partitioning across the
dipole for $\betaps = 0$ showers. In the \emph{split-dipole-frame} variant
(\pgzerosdf), we replace $f(\bar\eta) \to
f(\eta)$~\cite{FerrarioRavasio:2023kyg}. 

\subsection{PanLocal shower}
\label{subsec:ee-PL-Kinematic-map-inversion}
We next turn to the PanLocal map, as introduced in \cite{Dasgupta:2020fwr}.
\subsubsection{Forward map}
In contrast to the PanGlobal construction, the recoil associated with the branching is
redistributed locally within the emitting dipole. For a branching
$\{\tilde \imath,\tilde \jmath\}\to\{i,j,k\}$, with $\tilde \imath$ assigned as the
emitter and $\tilde \jmath$ as the spectator, the forward map is given by
\begin{subequations}\label{eq:panlocal-map}
  \begin{align}
    p_k &=
          a_k \tilde p_i
          + b_k \tilde p_j
          + k_\perp ,
          \label{eq:panlocal-map-k}
    \\
    p_i &=
          (1-a_k)\tilde p_i
          + \frac{a_k b_k}{1-a_k}\tilde p_j
          - k_\perp,
          \label{eq:panlocal-map-i}
    \\
    p_j &=
          \frac{1-a_k-b_k}{1-a_k}\tilde p_j .
          \label{eq:panlocal-map-j}
  \end{align}
\end{subequations}
The transverse momentum $k_\perp$ and the light-cone components $a_k$
and $b_k$ are defined as in
Eqs.~(\ref{eq:PG-kperp-def})--(\ref{eq:ak-bk-Sudakov-coeffs}). 
Note that in contrast to the PanGlobal map, the emitter $i$ and
spectator $j$ transform differently.
The choice of which dipole end is the emitter is made differently for
each of the two terms in the PanLocal analogue of Eq.~(\ref{eq:dPh}).

\subsubsection{Inversion}
Starting from the post-branching momenta and a given
assignment of the emitter, spectator and radiation, we determine the
corresponding pre-branching configuration and reconstruct the
splitting variables.
The light-cone components $a_k$ and $b_k$ are given by
\begin{equation}
a_{k}\,=\, \frac{s_{jk}}{s_{ij}+s_{jk}}\,,\qquad
b_{k}\,=\, \frac{s_{ik}}{s_{ij}+s_{jk}}\,\frac{s_{ij}}{d^{2}}\,,\label{eq:ee-PL-bk}
\end{equation}
where we continue to use the notation of Eqs.~(\ref{eq:PG-pd-pm-dsq})
and (\ref{eq:sxy}). 
The pre-branching scalar products involving the dipole momenta $\tilde
\imath$ and $\tilde \jmath$ are
\begin{subequations}\label{eq:ee-PL-tilde-sjQ}
\begin{align}
s_{\tilde{\imath}\tilde{\jmath}}\,=\, & d^{2}\,,\\
s_{\tilde{\imath}}\,=\, & s_{i}+s_{k}-\frac{s_{ik}}{s_{ij}+s_{jk}}\,s_{j}\,,\\
s_{\tilde{\jmath}}\,=\, & s_{j}+\frac{s_{ik}}{s_{ij}+s_{jk}}\,s_{j}\,.
\end{align}
\end{subequations}
The complete set of pre-branching scalar products involving an
arbitrary light-like momentum $\tilde {l}$ external to the emitting dipole
is given by 
\begin{subequations}\label{eq:ee-PL-tilde-sllprime}
\begin{align}
s_{\tilde{\imath}\tilde{l}}\,=\, & s_{il}+s_{kl}-\frac{s_{ik}}{s_{ij}+s_{jk}}\,s_{jl}\,,\\
s_{\tilde{\jmath}\tilde{l}}\,=\, & \frac{d^{2}}{s_{ij}+s_{jk}}\,s_{jl}\,,\\
s_{k_{\perp}\tilde{l}}\,=\, & -\frac{\text{Tr}[\slashed{i}\slashed{k}\slashed{j}\slashed{l}]}{s_{ij}+s_{jk}}+\frac{2\,s_{ik}\,s_{jk}\,s_{jl}}{\left(\,s_{ij}+s_{jk}\,\right)^{2}}\,,\\
s_{\tilde{l}}\,=\, & s_{l}\,,\\
s_{\tilde{l}\tilde{l}^{\prime}}\,=\, & s_{ll^{\prime}}\, , 
\end{align}
\end{subequations}
with $\text{Tr}[\slashed{a}\slashed{b}\slashed{c}\slashed{d}]=s_{ab}\,s_{cd}+s_{ad}\,s_{bc}-s_{ac}\,s_{bd}$.
As in the PanGlobal case, our matching procedure does not require 
the full set of four-momenta. Nevertheless, for completeness, we
provide them here:
\begin{subequations}
\begin{align}
\tilde{p}_{i}\,=\, & p_{i}+p_{k}-\frac{b_{k}}{1-a_{k}-b_{k}}\,p_{j}\,,\label{eq:ee-PL-tilde-pi}\\
\tilde{p}_{j}\,=\, & \frac{1-a_{k}}{1-a_{k}-b_{k}}\,p_{j}\,,\label{eq:ee-PL-tilde-pj}\\
k_{\perp}\,=\, & -a_{k}\,p_{i}-\frac{(1-2a_{k})\,b_{k}}{1-a_{k}-b_{k}}\,p_{j}+(1-a_{k})\,p_{k}\,.\label{eq:ee-PL-kperp}
\end{align}
\end{subequations}

\subsubsection{Jacobian}
The phase-space factorisation in the PanLocal shower holds similarly
to the PanGlobal case, i.e.\ 
works as in Eqs.~\eqref{eq:phsp-fac} and \eqref{eq:shower-rad-phsp} and is valid for 
an arbitrary number of emissions.
The Jacobian is fully 
determined by the pre-branching scalar products and branching variables
obtained by the inverse map. We find the Jacobian to be 
\begin{align}\label{eq:phsp-jac-pl}
  \mathcal{J}\,=\,s_{\tilde{\imath}\tilde{\jmath}}\,\frac{1-a_{k}-b_{k}}{\left(1-a_{k}\right)^{2}} \,.
\end{align}
The branching probability in the PanLocal shower is determined according to 
\eqref{eq:dPh}, using a different partition function, 
$f(\bar \eta) \to g(\bar \eta) $, as detailed in the supplemental 
material of \cite{Dasgupta:2020fwr}.
Recall that the two terms in square brackets in the analogue of
Eq.~(\ref{eq:dPh}) each correspond to a different choice of emitter in
the PanLocal kinematic map.
%

\section{Matching procedure}
\label{sec:match-proc}

Here, we will review how the ingredients presented in
Section~\ref{sec:shower-inversion} are used to implement nested
multiplicative matching, 
and discuss the associated challenges, notably those related to
coverage of phase space and boundedness of branching probabilities.
In particular we will examine how the methods from
Refs.~\cite{FerrarioRavasio:2023kyg,2026PLDoubleSoft} can be exploited
and extended to address these challenges.

\subsection{Multiplicative matching beyond one emission}
\label{sec:mult-matching-review}

In a parton shower, a given colour-ordered $n$-parton phase-space
point $\Phi_n$ can potentially be reached by more than one sequence of
branchings.
Consider for example the state $\bar q g_a g_b q$, with the particles
given in their colour-ordered sequence.
In an antenna shower, with a single kinematic map per dipole (e.g.\
Vincia or PanGlobal), that state might be reached either through
$\bar q q \to \bar q \tilde g_a q \to \bar q g_a g_b q$, with $g_b$
emitted from the $\tilde g_a q$ dipole, or through
$\bar q q \to \bar q \tilde g_b q \to \bar q g_a g_b q$, with $g_a$
emitted from the $\bar q \tilde g_b$ dipole.
Those two sequences are distinct ``histories''.\footnote{ In a dipole
  shower, like PanLocal, which has distinct kinematic maps for each
  dipole end, this example would have 8 potential histories for 
  obtaining the $\bar{q}ggq$ final state from a $\bar{q}q$ starting point.}

In a given history $h$, the shower step that leads to the state with
$i$ partons involves a splitting weight $\cS(b_i^h)$, differential in
the phase space.
Here $b_i^h$ denotes a set of branching variables that includes the
pre-branching phase-space point $\Phi_{i-1}^h$, the
shower ordering variable $v_i^h$, and further elements such as the
index of the dipole that splits, the rapidity, azimuthal angle, and flavour of the
emission.
Using the notation of Sec.~\ref{sec:shower-inversion}, specifically the
shower branching probability $d\cP/(d\lnv\, d\bar\eta \,d\phi)$, 
Eq.~(\ref{eq:dPh}), and the phase-space Jacobian $\cJ$,
Eqs.~(\ref{eq:phsp-jac-pg-lr}) and (\ref{eq:phsp-jac-pl}), we have 
\begin{equation}
  \cS(b_i^h) = \frac{d\cP}{d\lnv \, d\bar\eta \, \frac{d\phi}{2\pi}}
  \, \frac{16\pi^2}{2a_k b_k \cJ}\,,
\end{equation}
where we note the different meanings of $b_i^h$, the full set of branching
variables for the $i^\text{th}$ splitting, and of $b_k$, the quantity
entering the shower maps, Eqs.~(\ref{eq:PG-local}) and (\ref{eq:panlocal-map}).

Without matching, the shower's effective squared matrix element
$\cM_\text{shower}(\Phi_n)$ for a given (colour-ordered) $n$-parton ($n\ge 3$)
phase-space point $\Phi_n$ is given by a sum over all possible histories for
reaching that phase-space point
\begin{equation}
  \cM_\text{shower}(\Phi_n) = \sum_h  \cM^h_\text{shower}(\Phi_n).
\end{equation}
The effective matrix element $\cM^h_\text{shower}(\Phi_n)$ for
each history is given by
\begin{equation}
  \qquad
  \cM^h_\text{shower}(\Phi_n) =  \cM(\Phi_{2}^h)
  \prod_{i=3}^{n}
  \cS(b_i^h),
  \label{eq:history-weight-plain}
\end{equation}
where $\cM(\Phi_{2}^h)$ is the squared matrix element for producing
the original (Born) 2-parton state.

The essence of the matching approach is to multiply the branching
probabilities at each stage of the shower by an additional reweighting
factor $\cW(\Phi_i)$, chosen so as to ensure that the shower
reproduces the correct full squared matrix element $\cM(\Phi_{i})$ for
each step $i$.
Specifically one multiplies the standard shower branching probability
with a reweighting factor $\cW(\Phi_i)$
\begin{equation}
  \label{eq:PWithW}
 \cS_\text{matched} (b_i^h) = \cW(\Phi_i) \, \cS (b_i^h)\,,
\end{equation}
where
\begin{equation}
  \label{eq:Wdef}
  \cW(\Phi_i) =
  \frac{\cM(\Phi_i)}{\sum_h \cM^h_{\text{shower},\cW_{i-1}}(\Phi_i)}.
\end{equation}
Here the numerator has the true squared matrix element for the phase-space 
point, and the denominator involves the shower effective matrix
element, summed over all histories, including the reweighting factor
$\cW$ for all prior steps,
\begin{align}
  \cM^h_{\text{shower},\cW_{n-1}}(\Phi_n) =  \cM(\Phi_{2}^h)
  \left(
    \prod_{i=3}^{n-1}
    \cS_\text{matched} (b_i^h)
  \right)
  \cS (b_n^h)\,,
  \label{eq:history-reweighted-prior}
\end{align}
i.e.\ taking into account that steps before $n$ will already have been
reweighted.
Assuming $\cW$ is bounded to be below some value $\Omega$, we can then
generate branching $i$ with the splitting weight $\cS (b_i^h)$
multiplied by an enhancement factor $\Omega$, and accept the branching
with probability
\begin{equation}
  \label{eq:P-accept}
  P_\text{accept}(b_i^h) = \frac{\cW(\Phi_i)}{\Omega}\,.
\end{equation}
Nested multiplicative matching is attractive because in single-unresolved IR limits the
shower matrix element tends to the true matrix element, resulting in
$\cW = 1$, assuming proper accounting for spin
correlations~\cite{Collins:1987cp,Knowles:1988hu}. 
This means the corresponding singular limit of the shower is left unchanged, which
is valuable in terms of retaining logarithmic accuracy.

Multiplicative matching can, however, have certain critical difficulties.
Firstly, there may be phase-space points that cannot be reached by any
normal shower history.
For example with the PanGlobal shower with $\betaps=0$, we find that there is
about $20\%$ of the $n=4$ phase-space volume that cannot be reached
through ordered configurations.
\logbook{71689d0}{See analyses/test-ee-4jet-matrix-element-clean/res-logbook/ps-coverage-2d/test-ps-coverage-lnvbuffer0-ee4jet-paper.dat}
Reweighting will not make these accessible.
Secondly, even if all phase-space points are reachable, $\cW(\Phi_n)$
may not be bounded, making it difficult to use within a standard Monte
Carlo acceptance step.\footnote{Technically, accounting also for the
  Jacobian $\cJ$ between phase space and the shower's underlying
  random-number generation, it is sufficient for
  $\cW(\Phi_n) \, \cS\, a_k b_k \cJ$ to be bounded.
  This introduces additional degrees of flexibility, for example in
  the choice of the variable mapping from the random numbers.
  For the purposes of this article, however, we exploit this freedom
  only for the matching of the first emission, and just in the
  PanGlobal showers, as was done in Ref.~\cite{Hamilton:2023dwb}.  }
One approach to address these issues, adopted for example in
Vincia~\cite{El-Menoufi:2024sys,Altmann:2025yip}, is to add
additional direct $2 \to 4$ branchings (or, more generally $2 \to n$).
However, such an approach has yet to be explored in the context of
higher logarithmic accuracy.
Instead, we take the path that we and our
PanScales collaborators have used for implementing double-soft
corrections~\cite{FerrarioRavasio:2023kyg,2026PLDoubleSoft}. 
The two broad classes of method used there are (1) ``disordering'' of
emissions~\cite{2026PLDoubleSoft};
and (2) applying the acceptance probabilities to the sum of multiple
partonic configurations (``grouping'') and subsequently swapping
between configurations to achieve the correct relative weights
(``swaps'')~\cite{FerrarioRavasio:2023kyg}.
Below we review and extend these methods to the problem of
fully-matched 4-parton phase space.

\subsection{Disordering}
\label{sec:disordering}

Parton showers order emissions according to the value of some
``ordering variable'' $v$.
For example in the PanGlobal $\betaps=0$ showers, $v$ corresponds to
transverse momentum.
Even if repeated application of a shower's phase-space map can reach
any phase-space point, there may be points that cannot be reached from
an \emph{ordered} set of branchings.
This was recognised long
ago~\cite{ANDERSSON1990393,Andersson:1991he,Giele:2011cb}, however it
is only in the 
companion work~\cite{2026PLDoubleSoft} that it has become clear
how to exploit disordering while retaining logarithmic accuracy.

Specifically when a parton has been produced in step $i$ with some value
$v_i$ of the ordering variable, normally the shower would only allow
$\ln v_{i+1} < \ln v_i$.
Instead we allow the dipoles associated with that parton to start
showering from a scale $\ln v_{i+1} <  \ln v_i + \Delta$, where
$\Delta>0$ is of order $1$.
Allowing disordered emissions risks double counting certain phase-space regions.
Our key idea is to use disordering to account \emph{only} for
contributions that cannot be accessed through the normal ordered
showering.

Let us focus on the 4-parton matching case and assume the shower has been
matched up to three partons, i.e.\ we work with
$\cM_{\text{shower},\cW_3}^h(\Phi_4)$ from
Eq.~(\ref{eq:history-reweighted-prior}).
We sum the shower effective matrix elements separately over
ordered and disordered histories 
\begin{equation}
  \label{eq:w-ord-dis-def}
  s_\text{ord}
  = \sum_{h\in\text{ord}}
  \cM_{\text{shower},\cW_3}^h(\Phi_4),
  \qquad
  s_\text{dis}
  = \sum_{h\in\text{dis}}
  \cM_{\text{shower},\cW_3}^h(\Phi_4).
\end{equation}
Defining a further shorthand for the exact matrix element,
$e = \cM(\Phi_4)$, we partition it into two contributions, 
\begin{equation}
  \label{eq:wexact-partition}
  e = e_\text{ord} + e_\text{dis},  
\end{equation}
which are to be accounted for by the ordered and disordered
shower paths.
Specifically we take
\begin{equation}
  \label{eq:wexact-ord-dis}
  e_\text{ord} = \min(e,\, \Omega \mkern1.5mu s_\text{ord})\,,\qquad\quad
  e_\text{dis} = \max(e - \Omega \mkern1.5mu s_\text{ord},0)\,.
\end{equation}
When the shower is a good approximation to the exact matrix element,
the entirety of the exact matrix element is attributed to
$e_\text{ord}$, and disordered paths are simply rejected,
since $e_\text{dis}=0$.
The disordered contribution comes into play only if $\Omega \mkern1.5mu s_\text{ord}$
is an underestimate relative to the full matrix element.
That includes situations where the phase-space point is simply
inaccessible through an ordered path.

With the above scheme, we have separate acceptance probabilities
depending on whether the 4-parton phase-space point was reached
through an ordered or a disordered history
\begin{equation}
  P_{\text{ord}}
  =
    \frac{e_\text{ord}}{\Omega s_\text{ord}},
    \qquad
  P_\text{dis}
  =
    \frac{e_\text{dis}}{\Omega s_\text{dis}}.
\label{eq:pacc-both}
\end{equation}
Note that, by construction, the acceptance probability for the ordered
path is bounded, $0\le P_\text{ord}\le 1$.
However, while $P_\text{dis}$ is positive-definite, there is no
guarantee of an upper bound, a question that we address with the
aggregation and swap strategies of the next subsection.

The above algorithm involves two parameters: $\Delta$, which controls
how disordered two emissions may be, and $\Omega$, the
overhead.\footnote{In principle there also is freedom to choose
  distinct overhead factors for the ordered and disordered paths.}
We will use $\Delta=1$ and $\Omega = 2$ ($3$) for the PanGlobal
(PanLocal) case.

\subsection{Grouping and swaps}
\label{sec:grouping}

Eq.~(\ref{eq:Wdef}) and its generalisation with disordering are so far
implicitly for a specific colour-ordered phase-space point and
associated flavour structure.
The further idea that we exploit here is that rather than considering
Eq.~(\ref{eq:Wdef}) for a single phase-space point and flavour
structure at a time, we may consider groups of phase-space points and
flavour structures that are to be treated in common.

As an example, consider two states $\bar q g_a g_b q$ and
$\bar q q' \bar q' q$ (as before, given in a colour-ordered sequence).
Suppose that in a given phase-space region the shower generates too
little of $\bar q q' \bar q' q$, such that the reweighting acceptance
probability is above $1$. 
In contrast the shower tends to cover phase space more effectively for
the $\bar q g_a g_b q$ channel, because of its extra singularities.
In this case, we can choose to produce more than is needed of the
$\bar q g_a g_b q$ state, and then some of the time, relabel it as a
$\bar q q' \bar q' q$ state and so fill in the weight that was missing
for that channel.
For this to work, we therefore need to consider the $\bar q g_a g_b q$
and $\bar q q' \bar q' q$ channels grouped, rather than separately.
We can also choose groups of configurations defined through permutations of momenta.
For example, in work implementing the double-soft limits of
showers~\cite{FerrarioRavasio:2023kyg,2026PLDoubleSoft}, it was
convenient to consider the group of permutations of $\{g_a,g_b\}$ or
equivalently of $\{q' \bar q'\}$.\footnote{In that context it was
  referred to as a colour-flow swap, because swapping the $g_a, g_b$
  momenta in the colour-ordered $\bar q g_a g_b q$ configuration is
  equivalent to swapping between two colour flows.}

In what follows, we will outline two approaches to grouping and swaps.
The first of them is the approach used in
Refs.~\cite{FerrarioRavasio:2023kyg,2026PLDoubleSoft} and has the
property that swaps are \emph{always} active within a group.
This was sufficient for the PanGlobal class of showers.
The second is new to this article and activates the swaps only when
absolutely needed.
This was found to be useful for the PanLocal shower, specifically to
eliminate undesirable interplay between swaps and logarithmic accuracy
when many configurations are grouped together.

\subsubsection{Swaps that are always active}
\label{sec:swaps-always-active}

Suppose we have a group $g$ of configurations $i$, related through
some permutation of momenta and/or flavour.
When we reach a given shower configuration, we evaluate and sum the
effective shower matrix element not just for that configuration, but
all configurations $i$ in the same group, giving a shower effective matrix
element for the group as a whole, $s_g$, and similarly with the exact
matrix element, $e_g$
\begin{equation}
  \label{eq:sg-eg}
  s_g = \sum_{i\in g} s_i\, ,
  \qquad\quad
  e_g = \sum_{i\in g} e_i\,.
\end{equation}
We then apply an acceptance probability based on the group sums,
\begin{equation}
  \label{eq:group-accepated-prob}
  P_\text{\text{accept}} = \frac{e_g}{\Omega s_g}\,.
\end{equation}
This will ensure that the group as a whole has the correct
probability, but the relative weights between different permutations
$i$ within the group will, in general, be wrong.

The swap step performs this redistribution.
Define the normalised shower and exact channel fractions
\begin{align}
  \hat{s}_i = \frac{s_i}{s_g}, \qquad \hat{e}_i = \frac{e_i}{e_g}.
\end{align}
If the event has been generated in a channel $i$ for which $\hat{s}_i >
\hat{e}_i$, then the shower overpopulates that channel. With probability
\begin{align}
  \label{eq:swap-away}
  P_{i\to \text{other}} =
  \left(1 - \frac{\hat{e}_i}{\hat{s}_i}\right) \Theta(\hat{s}_i > \hat{e}_i)\,,
\end{align}
we swap such events out of permutation $i$.
The destination permutation is chosen among the underpopulated
permutations, $\hat{s}_j < \hat{e}_j$, with probability
\begin{align}
  \label{eq:swap-into}
  P_{i\to j | i\to \text{other}} =
  \frac{(\hat{e}_j-\hat{s}_j)\Theta(\hat{s}_j < \hat{e}_j)}
        {\sum_{k\in g} (\hat{e}_k-\hat{s}_k)\Theta(\hat{s}_k < \hat{e}_k)} .
\end{align}
Channels that are underpopulated do not swap away, they only
receive events from overpopulated channels.
Conversely, overpopulated channels only swap away.
The total accepted rate remains unchanged, while the
relative channel populations are adjusted from $\hat{s}_i$ to
$\hat{e}_i$.

This always-active swap scheme is what we adopt for the PanGlobal
shower.
In particular, we find that together with disordering, and
permutations of momenta and flavours of $\{g_a,g_b\}$ and
$\{q' \bar q'\}$ (the ``central'' particles), this scheme is sufficient to have
bounded acceptances, and is compatible with the double-soft IR limits.
The always-active swap scheme can be adapted to work in two stages, first performing
flavour swaps, followed by momentum swaps.
This is the strategy
adopted in Ref.~\cite{FerrarioRavasio:2023kyg} to correct for
double-soft emissions, and is also what we do here.

\subsubsection{Minimal swaps}
\label{sec:swaps-minimal}

When studying the PanLocal shower, we found that swaps of just the two
central particles were insufficient to cover phase space, even with
disordering.%
\logbook{0e93823f6c}{See
  logbook/2025-01-17-ee-4jet-me/phsp-coverage/coverage-overhead-checks-PL05/nperms+2-lnvbuff1-overhead3.dat,
  specifically the ``zero weight'' entries. Also the files with
  buffer2, buffer3, which increase the buffer, }
This led us to consider more ample sets of permutations.
In particular, to obtain not just phase-space coverage but also
bounded acceptances, we found it necessary to allow for all
permutations of the four momenta (modulo a reversal symmetry).
This led us to investigate alternative swap schemes.
In an ideal situation, the always-active swap scheme and the minimal
swap scheme (to be defined below) are both applicable.
However, if there are any single-unresolved limits with a mismatch
between the effective shower matrix element and the exact QCD matrix
element, the always-active swap scheme can result in long-distance
swaps (in a Lund-diagram sense) between distinct topologies whose
Sudakov form factors differ at the single-logarithmic level, hence a
breaking of logarithmic accuracy.
In the case of the PanLocal shower, this is triggered by a combination
of two factors: the large number of permutations and the fact that,
for the simplicity of our implementation, the effective shower matrix
elements do not explicitly include spin-correlation effects.
Appendix~\ref{sec:issues-with-pl-12-swaps} illustrates how this would
break logarithmic accuracy with the always-active swap scheme.

The minimal swap scheme takes the approach that if
a given permutation $i$ can be obtained within the overhead factor
without swaps, i.e.\ $e_i < \Omega s_i$, then one should never swap
into that permutation.
Swaps into another permutation are instead reserved for those cases where
the permutation has  $e_i > \Omega s_i$.
Of course we still need some swaps, to address non-singular limits
where $e_i > \Omega s_i$.
Specifically for each permutation $i$, we work out the ``headroom''
$\cH_i$ associated with the permutation,
\begin{equation}
  \label{eq:headroom}
  \cH_i = \Omega s_i - e_i\,.
\end{equation}
We break the set of permutations into two groups, $g_+$ and $g_-$
according to whether $\cH_i >0$ or $<0$.
Swaps are only carried out from a permutation $i$ with $\cH_i>0$ into
a permutation $j$ with $\cH_j < 0$, with a probability proportional
to $\cH_i$ and $|\cH_j|$.
The acceptance probability for permutations with
$\cH_i < 0$ is simply $P_{\text{accept},i}=1$.
The acceptance and swapping probabilities for permutations in the
$g_+$ group can be determined from the constraints that (a) their
expected QCD weight is reproduced, and (b) that the total deficit,
$\cH_-=\sum_{j\in g_-}\cH_j$, must be filled in by swaps
from $g_+$ (proportionally to $\cH_i$).
This means
\begin{equation}
  P_{\text{accept},i} (1-P_{i \to \text{other}}) = \frac{e_i}{\Omega s_i},
  \qquad
  P_{\text{accept},i} P_{i \to \text{other}} = \frac{1}{\Omega s_i} \frac{|\cH_-|}{\cH_+}\cH_i,
  \qquad
  \text{for $i \in g_+$},
\end{equation}
with $\cH_\pm=\sum_{j\in g_\pm}\cH_j$.
We deduce
\begin{equation}\label{eq:Pacc-pos-headroom}
  P_{\text{accept},i} = \frac{1}{\Omega s_i}
  \left(
    e_i + \frac{|\cH_-|}{\cH_+}\cH_i
  \right),
  \qquad
  P_{i \to \text{other}} = \frac{
    \frac{|\cH_-|}{\cH_+}\cH_i
  }{
    e_i +  \frac{|\cH_-|}{\cH_+} \cH_i
  },
  \qquad
  \text{for $i \in g_+$}.
\end{equation}
The choice of destination $j$ is then given by the probabilities             
\begin{equation}
  \label{eq:minimal-P-i-to-j}
  P_{i \to j | i \to \text{other}} =
  \frac{\cH_j}{\cH_-},
  \qquad
  \text{for $i \in g_+$, $j \in g_-$}.
\end{equation}
In practice, we apply this algorithm just for the momentum
permutations, where in each momentum permutation we sum over flavour
permutations.
We still use disordering, and effectively replace $e$ in
Eq.~(\ref{eq:wexact-ord-dis}) with $e_i+(|\cH_-|/\cH_+)\cH_i$ for
channels with positive headroom, or with $\Omega s_i$ for channels with
negative headroom, and in all formulas of this section $s$ is to be
understood as the sum of ordered and disordered
histories.\footnote{Note that the procedure requires $|\cH_-| \leq
  \cH_+$, and so $e_g \leq \Omega s_g$ for the entire group.} 
We note that channels corresponding to single-unresolved limits will always
have positive headroom, as long as $\Omega$ is sufficiently large to
account for any spin correlation missing in the shower effective
matrix element.%
\footnote{In the $\gamma^*/Z \to 4$-parton case, the singular
  limit of the spin-correlated matrix element, i.e.\ the
    problematic case for the always-active scheme with the PanLocal shower, see
    Appendix~\ref{sec:issues-with-pl-12-swaps}, is at most twice
  the spin-averaged result and so with $\Omega \ge 2$ we will always
  have $e_i < \Omega s_i$.  }
One can easily show that, in that case, the
ordered path alone is sufficient to reproduce the full exact QCD
matrix element, with the disordered contribution vanishing, thus
preserving logarithmic accuracy in these limits.
After the momentum permutation has been decided, we then use the
always-active algorithm for swapping between flavour channels.

\section{Validation}
\label{sec:validation}

We have carried out a variety of tests to verify the correctness of
our matching procedures.
Some of them can be considered ``technical'': for example we have
generated a large number of phase-space points ($10^{11}$ for the
PanLocal case)%
\logbook{c06d8cc9b}{logbook/2025-01-17-ee-4jet-me/phsp-coverage/coverage-overhead-checks-PL05/nperms+12-lnvbuff1-overhead3-allrseq.dat}
with the Rambo
algorithm~\cite{Kleiss:1985gy} and verified that there always exist
ordered or disordered shower paths that reach each point, with the
disordering buffer $\Delta=1$.
We have numerically integrated the 4-parton phase-space volume with the
Jacobians of Section~\ref{sec:shower-inversion} to verify that it
comes out correctly.
We have also checked that the acceptance probabilities are always
bounded to be less than $1$.
And we have verified that as one approaches singular limits, the swap
rate either tends to zero (in single-unresolved cases, see
Appendix~\ref{sec:issues-with-pl-12-swaps}), or remains active only
within an 
appropriate singular subset of particles (e.g.\ a triple-collinear
or double-soft subset).
\logbook{cd0f01bc57}{this is delicate especially in the 12-swap case, see
  analyses/test-ee-4jet-matrix-element-clean/2026-06-pl-swap-rates pl-swap-rates.pdf}

In this section, we focus on tests involving specific observables.
We first check, in Section~\ref{sec:fo-val}, that the matched shower
reproduces fixed-order predictions for a set of four-jet event-shape
observables after two emissions.
In Section~\ref{sec:DU-limits}, we
examine the behaviour of the matching in double-unresolved regions,
focusing in particular on the double-soft limit,
which is relevant for the higher logarithmic accuracy of the
shower.
Finally, in Section~\ref{sec:log-accuracy-tests}, we present
all-order logarithmic tests to verify that the matching procedure does
not spoil the logarithmic accuracy of the shower.

Note that the matching and the corresponding comparisons are all
carried out at leading $\nc$.

\subsection{Fixed-order tests}
\label{sec:fo-val}
We examine several 4-jet observables: the $D$-parameter~\cite{Parisi:1978eg,Donoghue:1979vi,Ellis:1980wv}, the
Thrust minor~\cite{Brandt:1964sa,Farhi:1977sg}, the light-jet mass~\cite{Clavelli:1979md} and $y_{34}$ of the Durham
algorithm~\cite{Catani:1991hj}.
Their distributions all start at order $\as^2$ and we compare the
distributions as obtained with the showers to results from the
\texttt{Event2} program~\cite{Catani:1996jh,Catani:1996vz}.

The PanScales code can operate in two modes: it can generate pure
tree-level events without a Sudakov form factor, or it can be run as a
full shower.
The implementations of the two modes overlap considerably, but there
are some points of difference.
Accordingly we carry out tests in both modes.

\begin{figure}
  \centering
  \includegraphics[width=0.48\textwidth,page=1]{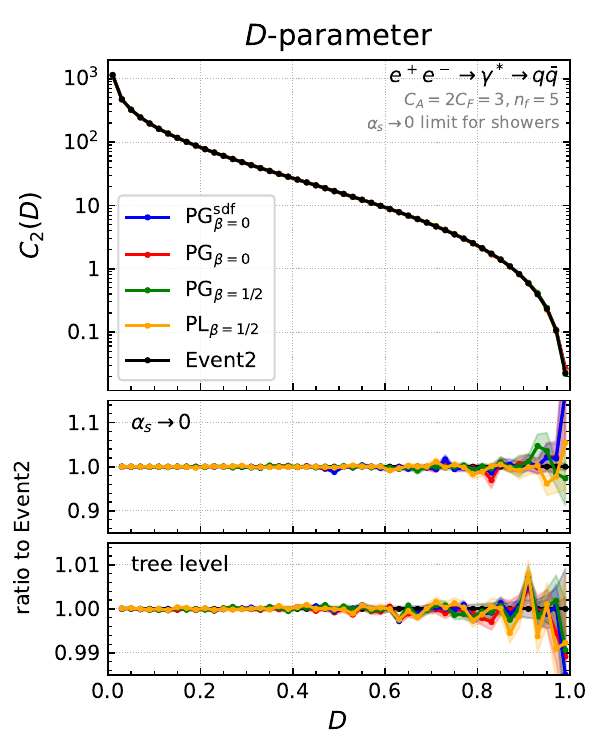}
  \includegraphics[width=0.48\textwidth,page=2]{figures/plot_EventShapes_3panels.pdf}
  \includegraphics[width=0.48\textwidth,page=3]{figures/plot_EventShapes_3panels.pdf}
  \includegraphics[width=0.48\textwidth,page=4]{figures/plot_EventShapes_3panels.pdf}  
  \caption{Comparison of fixed-order results
    (\ttt{Event2}) with PanGlobal
    (PG) and PanLocal (PL) showers with different ordering choices
    $\beta\equiv\betaps$.
    The comparison is performed for various event shapes,
    namely $D$-parameter, $y_{34}$ with Durham algorithm, Thrust
    minor and light-jet mass.
    In the top panel, we run the full shower until the point where it
    reaches the infrared cutoff or has generated two emissions, and we
    do so for multiple values of $\as$ and then extrapolate to $\as=0$ as in
    Eq.~(\ref{eq:dsigma-dV-div-as2}) to obtain the pure $\as^2$ coefficient.
    The middle panel shows the ratio of that to \texttt{Event2}
    results.
    The bottom panel shows the ratio of a pure tree-level run of the matched
    shower code to \texttt{Event2}.
    Note the much smaller vertical scale.
  }
  \label{fig:bp2-fo-test}
\end{figure}

In shower mode, there are contributions beyond $\as^2$, e.g.\ from the
Sudakov form factor, the running of the coupling, etc.
We carry out the standard PanScales $\as \to 0$ type
extrapolation to extract the coefficient of $\as^2$ for the distribution.
\begin{equation}
  \label{eq:dsigma-dV-div-as2}
  C_2(V) = \lim_{\as \to 0} \frac{1}{\as^2} \frac{V}{\sigma_0} \frac{d\sigma}{dV}\,,
\end{equation}
where $\sigma_0$ is the Born cross section.
To limit the computational intensity of the test, we stop the shower
after it has generated two emissions, since further emissions will
only modify terms beyond order $\as^2$.
The values of $\as$ that we use are $\as = \{0.1, 0.01, 0.005,
0.001$\}.

The results in shower mode are shown in the top panel of Fig.~\ref{fig:bp2-fo-test},
with the ratio to \texttt{Event2} given in the middle panel.
One sees good agreement to within the statistical errors for each of
the four observables and for all four showers that we test.
It is important to keep in mind that even within a given class of
shower, e.g.\ PanGlobal, the relative amount of disordering differs,
for example according to the choice of shower-ordering variable $v =
k_t e^{-\betaps |\eta|}$.
Furthermore, the set of observables we consider also has
different scalings with the rapidity $\eta$ of soft emissions. The
overall agreement across these four showers and four observables is
therefore a strong validation of our matching procedure.

The bottom panels of Fig.~\ref{fig:bp2-fo-test} show the ratio of pure
tree-level runs of the matched PanScales code.
These are computationally less expensive and so allow for tests at
higher accuracy, as can be seen from the vertical scale.
Again we see good agreement to within the statistical errors, which
are below per mil level in much of the distribution.

\begin{figure}
  \centering
  \includegraphics[width=0.50\textwidth,page=1]{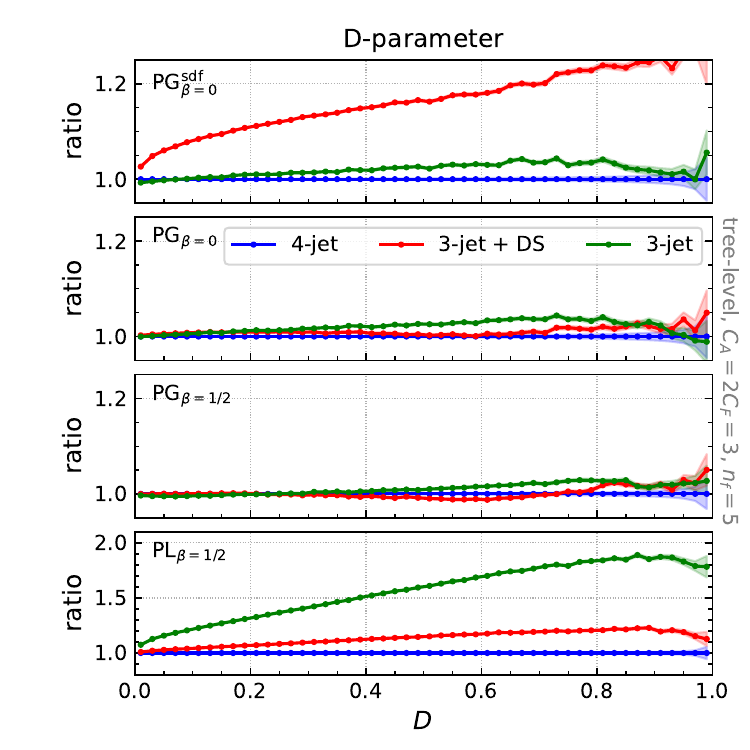}%
  \includegraphics[width=0.50\textwidth,page=2]{figures/BP2-vs-BP1.pdf}\\
  \caption{Illustration of deviation from exact leading-colour 4-jet
    result when running the showers without 4-parton matching, at tree-level.
    The deviation is shown with \emph{just} $3$-jet matching and with
    $3$-jet matching plus double-soft (DS) corrections.
    Each panel is for a different shower and/or $\beta_\ps$.
    Note the much larger scale for the \plhalf panel.
    The left-hand (right-hand) plot is for the $D$-parameter (Durham
    $y_{34}$ clustering scale). 
  }
  \label{fig:tree2-no-bp2}
\end{figure}

To give the reader a sense of the impact of the 4-parton matching,
Fig.~\ref{fig:tree2-no-bp2} shows the ratio of showers with just
$3$-parton matching to the exact leading-colour tree-level
$D$-parameter distribution (left) and the Durham $y_{34}$ distribution
(right).
Let us first focus on the green curves, where double-soft matching is
turned off. 
It is intriguing that the PanGlobal showers without double-soft
corrections are within about ten percent (or even better) of the correct answer.
We do not have any understanding as to why.
In contrast the PanLocal shower is wrong by up to a factor of two,
which is roughly what one might expect given the absence of
matrix-element corrections.
Once double-soft corrections are turned on (red curves), the
PanLocal shower moves substantially closer to the correct result, to
within about $20\%$ of it across the spectrum, suggesting that the
double-soft corrections embody a significant part of the true
$4$-parton matrix elements.
In a sense this is what one might have hoped for: as singular limits
are constrained, there is also less freedom away from the singular
limits.
Also of note is that for one of the PanGlobal variants
(PG$_{\beta=0}^\text{sdf}$), the double-soft corrections move it away
from the correct result.
Nevertheless it is still within $\sim25\%$  of the correct result.

The event-shape tests of Fig.~\ref{fig:bp2-fo-test} are inclusive over
flavour and over momentum permutations.
We have also checked the results agree with \texttt{Event2} separately
in the $\bar q gg q$ and $\bar q q' \bar q' q$ channels.
The correctness of the momentum permutations will be explicitly
displayed in
the tests in the next section.
A final comment here though is that if one examines the analogues of
Figs.~\ref{fig:bp2-fo-test} and \ref{fig:tree2-no-bp2} for specific flavour
channels, one finds that the relatively small 4-quark channel is
significantly less well approximated by the shower, even with
double-soft corrections turned on.
This suggests that 4-parton matching might be especially important for
secondary flavour production studies.

\subsection{Consistency of double-unresolved limits}
\label{sec:DU-limits}

\begin{figure}[t!]
  \centering
  \includegraphics[width=0.75\textwidth,page=3]{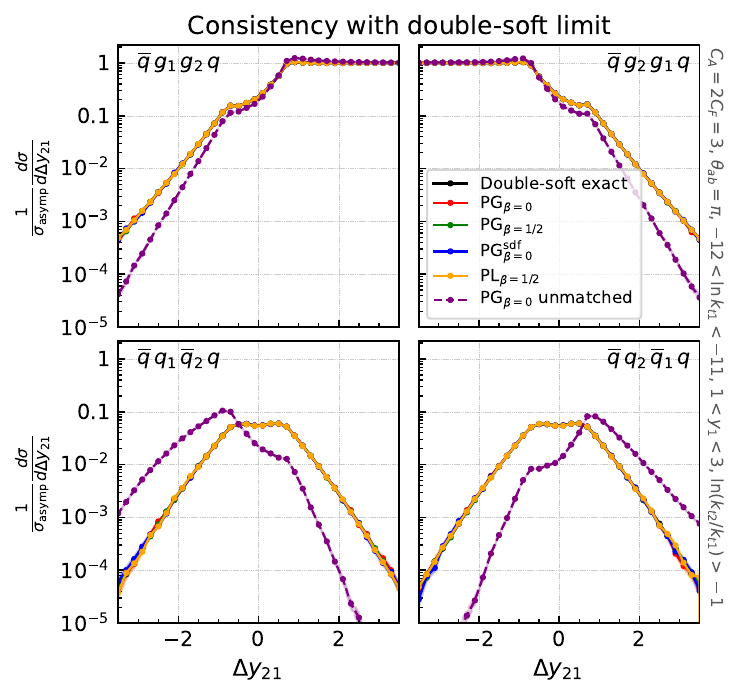}
  \caption{Colour-flow and flavour-channel decomposition of several
    showers with 4-parton matching, in the double-soft limit, as well
    as one unmatched shower, for comparison.
    The plot shows the distribution of $\Delta y_{21} = y_2 - y_1$ in
    a window $\ln k_{t2}/k_{t1} > -1$, where the rapidities $y_i$ and
    transverse momenta $k_{ti}$ are obtained from Lund declustering.
    Labels such as $\bar q g_1 g_2 q$ indicate the colour flow or
    flavour assignment in the numbering where $k_{t1} > k_{t2}$. 
    We use the convention that $\bar q$ ($q$) is at negative (positive) 
    rapidities.
  }
  \label{fig:bp2-fo-ds-channels}
\end{figure}

As a first check, we test the 4-parton matching in the double-soft
limit.
This is a direct analogue of the test in the supplementary material
of Ref.~\cite{FerrarioRavasio:2023kyg}.
It isolates a double-soft configuration by starting from a hard dipole
$ab$ and generating two emissions that are both soft with respect to
the initial hard partons.
In practice we carry out a Lund declustering~\cite{Dreyer:2018nbf} of
the event, and then keep the event if the emission with larger
transverse momentum, labelled $1$, is in a given window of
soft transverse momentum, $-12 < \ln k_{t1}/Q < -11$, and relatively
central rapidity $1 <y_1 < 3$, and the second emission is within one
logarithmic unit in transverse momentum, $\ln k_{t2}/k_{t1}>-1$.
Fig.~\ref{fig:bp2-fo-ds-channels} shows the distribution of
$\Delta y_{21}=y_2-y_1$ for several showers with 4-parton matching, as
well as the expectation based on the double-soft
limit~\cite{Dokshitzer:1992ip,Campbell:1997hg,Catani:1999ss} of the
matrix element.
We have taken the ratio between the differential distribution in
$\Delta y_{21}$ and the leading-colour expectation for asymptotically
large rapidity separations, i.e.\ $y_2 - y_1 \gg 1$, which reads
\begin{align}
  \mathrm{\sigma_\text{asymp}}= \sigma_0 \spac \left(\frac{\as \spac \nc}{\pi}\right)^2 \times 2 \spac ,
\end{align}
for the specific cuts we are applying (i.e.\ taking into account
$\Delta\ln k_{t,1}=1$, $\Delta y_1=2$).
The plot shows good agreement of the shower with the exact double-soft
result, and confirms also that the showers are correctly
providing the two separate colour flows, as well as the quark--gluon
separation.
\begin{figure}
  \centering
  \includegraphics[width=0.85\textwidth]{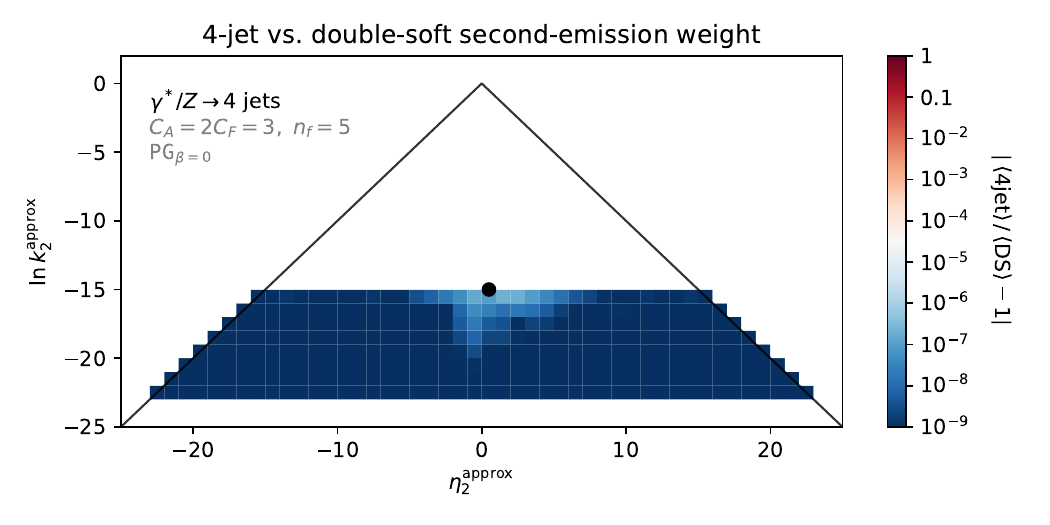}
  \caption{Absolute value of the relative  difference in average shower weights between the 4-parton
    matching and the double-soft generation of a second soft parton,
    after manual injection of a first soft parton.
    The difference is shown as a function of the rapidity and
    transverse momentum of the second emission.
    The first emission is fixed at the location of the black
    disc.
    Note that for simplicity, the test is shown only with
    \emph{ordered} events. 
    The Lund coordinates are based on an internal shower routine that
    uses a soft-collinear approximation~\cite{Hamilton:2020rcu}.
    The secondary Lund plane has been ``folded'' onto the primary Lund
    plane, i.e.\ a given bin of rapidity and transverse momentum may
    include entries from both the primary and secondary Lund planes.}
  \label{fig:DS-point-by-point}
\end{figure}

Our next test relates to the discussion of
Section~\ref{sec:matching-log-accuracy} and focuses on
configurations like those of Fig.~\ref{fig:ds-avoid-SLLs}.
Specifically, we consider events where we manually inject a first soft
large-angle emission, and then generate a second emission.
We have two sets of events, one where the second emission is generated
by the shower's double-soft code and another where it is generated by
the 4-parton matching code.
We do this in the ``tree-level'' variant of the PanScales code, mainly
because this facilitates alignment of random numbers exactly between
the double-soft and the 4-parton paths.
We identify the Lund coordinates of the 2nd emission, and then in each
bin of the Lund plane we take the relative difference between the
average bin weights of the double-soft and 4-parton paths.
That is shown for the PanGlobal ($\betaps=0$) shower in
Fig.~\ref{fig:DS-point-by-point}, with the location of the first
emission represented with a black disc.
One sees that the average difference is numerically small and, in the
double-soft region, consistent with being of order $k_t/Q$, as one
would have expected from the neglect of the hard-parton recoils in the
double-soft approximation.
Repeating the test with a yet lower scale for the injection further
substantially reduces the difference.
This provides a powerful verification that the double-soft and 4-parton
codes align in the double-soft region.
\logbook{f6f53f8}{Verified also for PG05; explicit verification for
  PL05 was technically more complicated, not done. See
  analyses/test-ee-4jet-matrix-element-clean/res_DS_vs_4jet_injection/
  the data, run scripts and the pdfs are there}
The level of agreement between the binned conditional weights here is
consistent with a \emph{vanishing} leading-power mismatch coefficient
$C$ in Eq.~(\ref{eq:sll-issue}), providing a numerical check of the double-soft
alignment needed to avoid the spurious SLL generating mechanism
illustrated in Fig.~\ref{fig:ds-avoid-SLLs}. 

Note that were we to repeat the test with an energetic collinear first
emission, we would see $\order{1}$ discrepancies, localised in the
triple-collinear region.
This is simply because the shower does not yet contain the
triple-collinear matrix element and connects with the caveat in the
last paragraph of Section~\ref{sec:matching-log-accuracy}.

\subsection{All-order tests}
\label{sec:log-accuracy-tests}

The tests of the previous subsections have shown the tree-level
correctness of the overall 4-parton matching and its consistency with
the implementation of the double-soft matrix element in the PanScales
showers.

For consistency with NNLL logarithmic accuracy for event shapes,
non-global logarithms and multiplicities, it is not just the real
matrix element that needs to be handled but also the loop corrections.
We do not currently have the full loop corrections for matching, e.g.\
3-jet at NLO, however it is important that in the soft and collinear
limits we retain consistency with the loop corrections of
Refs.~\cite{FerrarioRavasio:2023kyg,vanBeekveld:2024wws}.
We do this by applying the relative $\as$ and $\as^2$ corrections from
those references to the rate, not just of normal shower
splittings, but also of the splittings used in 4-jet
matching.\footnote{With the proviso that for the PanLocal shower, the
  double-soft correction has been updated to use the ``minimal'' swap
  scheme of Section~\ref{sec:grouping}, with a corresponding
  re-calculation of NNLL corrections to splitting rates.}
This ensures that when the 4-parton matching is being applied in
singular limits, the splitting rates remain consistent with the needs of
logarithmic accuracy.

\begin{figure}[t!]
  \centering
  \includegraphics[width=1.0\textwidth,page=1]{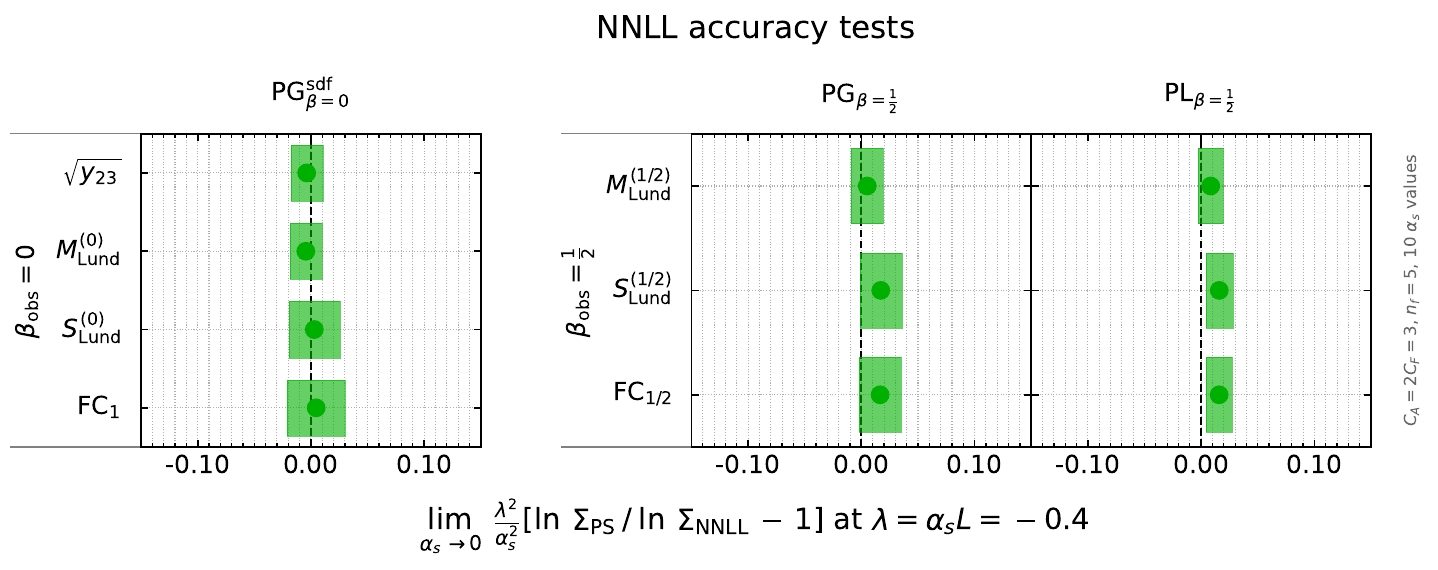}
  \caption{Summary of the results for log-accuracy tests for event
    shapes, with $\betaobs=\betaps$.
    The left-hand panel shows $\betaps=0$ tests for the PanGlobal
    shower, while the right-hand panel shows $\betaps=0.5$ tests for
    the PanGlobal and PanLocal showers.
    The observables are as defined in
    Refs.~\cite{Dasgupta:2020fwr,vanBeekveld:2025zjh} 
    and \cite{Banfi:2004yd}.
  }
  \label{fig:bo2-nnll-summary}
\end{figure}

Full logarithmic accuracy tests are computationally expensive.
Accordingly here we carry out only a subset of them.
Specifically we consider event shapes with $\betaobs =
\betaps$, where we recall that a given event shape and the shower
ordering variable have a scaling $k_t e^{-\beta |\eta|}$.
We apply the standard PanScales procedure~\cite{Dasgupta:2020fwr},
performing a limit for $\as\to 0$, while $\as L=\lambda$ is kept
fixed, where $L$ is the large logarithm associated with the
observable. More specifically, we consider the NNLL residual
\begin{align}
  \Delta_\text{NNLL}(\lambda) = \lim_{\as\to 0}
  \left(\frac{\lambda}{\as}\right)^2 \left[ \frac{\ln\Sigma_{\rm
        PS}(\lambda)}{\ln\Sigma_\text{NNLL}(\lambda)} - 1 \right] ,
  \label{eq:nnll-residual}
\end{align}
where $\Sigma_\text{PS}$ is the cumulative distribution obtained from
the matched shower and $\Sigma_\text{NNLL}$ is the corresponding
analytic NNLL prediction~\cite{Banfi:2016zlc,Banfi:2018mcq,vanBeekveld:2025zjh}.
Summary results are shown in Fig.~\ref{fig:bo2-nnll-summary},
demonstrating good agreement to within statistical uncertainties.

\begin{figure}
  \centering
  \includegraphics[width=0.9\textwidth]{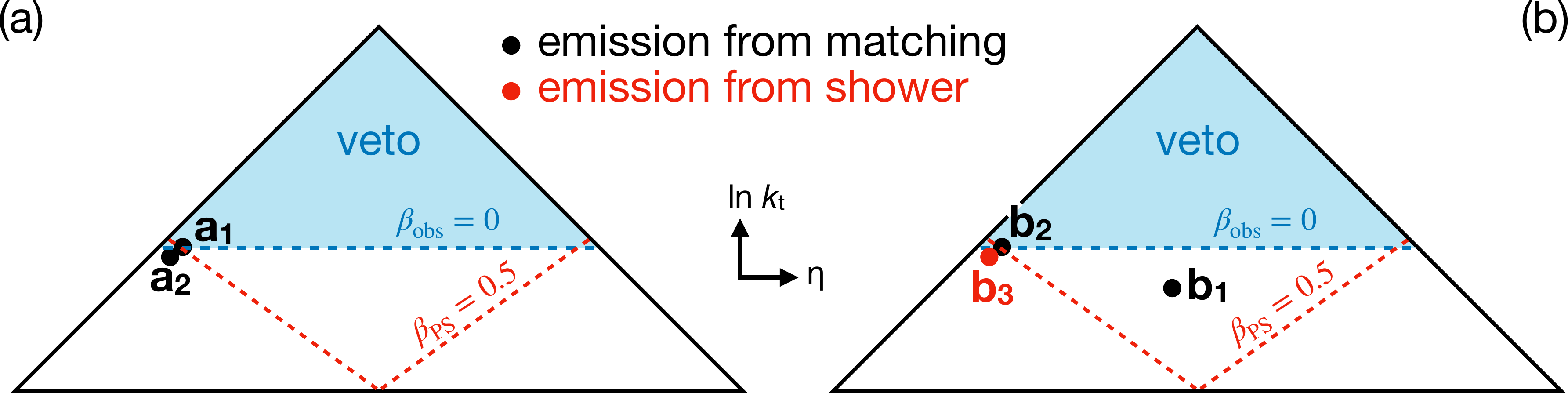}
  \caption{Illustration of the issue that arises for
    $\betaps > \betaobs$ due to the absence of triple-collinear
    splitting in the shower. See text for details.}
  \label{fig:tc-evshp-issue}
\end{figure}

One important point relates to the caveat from the last
paragraph of Section~\ref{sec:matching-log-accuracy}.
Specifically, since the showers do not yet include the
triple-collinear splitting functions, we have to watch out for terms
analogous to those of Eq.~(\ref{eq:sll-issue}), which in this context
would violate exponentiation of the logarithmic series.
This is illustrated in Fig.~\ref{fig:tc-evshp-issue}.
It shows the observable contour for $\betaobs=0$ in blue, together
with the blue-shaded vetoed region, as well as a $\betaps=0.5$ shower
contour in red.
On the left-hand side, two emissions in the triple-collinear region
($a_1,a_2$) are produced by the matching and will have the correct
triple-collinear limit.
On the right-hand side, one of the matched emissions, $b_1$, was
produced well below the region vetoed by the observable, but much earlier
in the shower history than $b_2$ and $b_3$.
It has no impact on the observable, but has the result that of the two
emissions in the triple-collinear region ($b_2,b_3$), one is produced
by the shower and so does not have the correct triple-collinear
distribution.
The relative weights of the left- and right-hand side events involve
the same kind of dependence on an $e^{-\as L^2}$ factor as in
Eq.~(\ref{eq:sll-issue}), which would break exponentiation (as per
discussion also in Ref.~\cite{Dasgupta:2018nvj}).
Note that the issue does not arise if we have $\betaps \le \betaobs$,
e.g.\ as in Fig.~\ref{fig:bo2-nnll-summary}, because the
shower reaches the triple-collinear matching region \emph{before} any
double-log non-vetoed region.
On the other hand, more generally, it \emph{would} be relevant for
purely collinear quantities such as fragmentation functions.
%
\section{Conclusions}
\label{sec:conclusions}

One of the key observations of this paper is the importance of the
principles outlined in Section~\ref{sec:matching-log-accuracy}.
Specifically, when interfacing a hardest-emission generator
(HEG) with a parton shower, it is important that the
HEG and the shower should have identical emission probabilities in the
infrared limit, for as many emissions as provided by the HEG.
Even just a small, high-order mismatch can cause exponentiation to
break for event-shape type observables, and generate spurious
super-leading logarithms for single-logarithmic observables.

The approach that we took to achieve (leading-colour) 4-parton
matching was to reweight the parton showers themselves outside the
singular limits: a form of nested multiplicative matching.
To do this, we established how to invert an arbitrary number of steps
of the PanGlobal and PanLocal showers, and computed the phase-space
Jacobians for any multiplicity.
For up to two emissions, we achieved full phase-space coverage and
bounded weights with the help of the techniques originally developed
for the double-soft limit of showers: disordered emissions as in
Ref.~\cite{2026PLDoubleSoft}, and an extension of the grouping and
swaps of momentum configurations from
Ref.~\cite{FerrarioRavasio:2023kyg}.

The resulting full 4-parton reweighting agrees with results
from the \ttt{Event2} program~\cite{Catani:1996jh,Catani:1996vz},
Fig.~\ref{fig:bp2-fo-test}, within statistical uncertainties, $\ll 1\%$.
For the PanGlobal shower the matching corrections are a modest effect
(less than $25\%$ for common observables), while they are large, up to
$90\%$, for the PanLocal shower, Fig.~\ref{fig:tree2-no-bp2}.
Interestingly the PanLocal corrections are substantially reduced if
one considers them relative to a 3-parton plus double-soft matched
baseline, as if some key features of the hard region were effectively
already encoded through the double-soft limit.

We also verified that our approach reproduces the double-soft
reweighting of Refs.~\cite{FerrarioRavasio:2023kyg,2026PLDoubleSoft}
in the corresponding limit.
We carried out the test both in an averaged sense
Fig.~\ref{fig:bp2-fo-ds-channels}, and bin-by-bin, cf.\
Fig.~\ref{fig:DS-point-by-point}, with the latter providing the
guarantee of the principles of
Section~\ref{sec:matching-log-accuracy} in the double-soft limit,
though not yet in the triple-collinear limit, because this is not
implemented in the shower.
Finally we carried out tests that we retain NNLL accuracy for common
event-shape distributions, Fig.~\ref{fig:bo2-nnll-summary}, modulo the
question of exponentiation associated with the lack of
triple-collinear shower terms, as discussed around
Fig.~\ref{fig:tc-evshp-issue}.

We defer the study of the phenomenology to separate
work~\cite{PanPheno} that will bring together the double-soft
limit and NNLL terms across multiple showers~\cite{2026PLDoubleSoft},
as well as the 4-parton matching performed here. 

Overall, the developments presented here represent an important step on
the path to logarithmically consistent NNLO matching.

\section*{Acknowledgements}

We are grateful to our PanScales collaborators (Sam Bates, Mrinal
Dasgupta, Basem El-Menoufi, Silvia Ferrario Ravasio, Jack Helliwell,
Petr Jakubcik, Alexander Karlberg, Matvei
Kuzmin, Marco Leitão, Pier Monni,  Ludovic Scyboz, Alba Soto-Ontoso)
for their work on the code and the underlying 
philosophy of the approach, and particularly wish to thank Alexander
Karlberg and Silvia Ferrario Ravasio for their comments on this manuscript.
This research was supported
by the Dutch Research Council (NWO) under project number
VI.Veni.232.190 (MvB),
by the French Agence Nationale de la Recherche (grant
ANR-25-CE31-7192, P2S2) (GS),
by the Science and Technology Facilities Council (STFC)
under grants ST/X000761/1 (GPS, SZ), UKRI4236 (GPS),
UKRI4244 and ST/X000516/1 (KH),
and by a Royal Society Research Professorship
(RP$\backslash$R1$\backslash$231001) (GPS, NS).

The authors would also like to acknowledge the use of the University
of Oxford Advanced Research Computing (ARC) facility in carrying out
part of this work, \url{https://doi.org/10.5281/zenodo.22558}, as well
as the Oxford Theoretical Physics Hydra cluster, the Stoomboot cluster
at Nikhef, and the Wilson cluster at the ITP in Bern.

\appendix
\section{Spin correlations and minimal swaps}
\label{sec:issues-with-pl-12-swaps}

Here we illustrate the issue that forces us to use the minimal swap
scheme for the PanLocal shower, rather than the always-active scheme.
Specifically, we consider what happens with the always-active scheme,
with swaps between 2 flavour permutations, followed by swaps between
12 momentum permutations.

\begin{figure}[p]
  \centering
  \fbox{\includegraphics[width=0.9\textwidth]{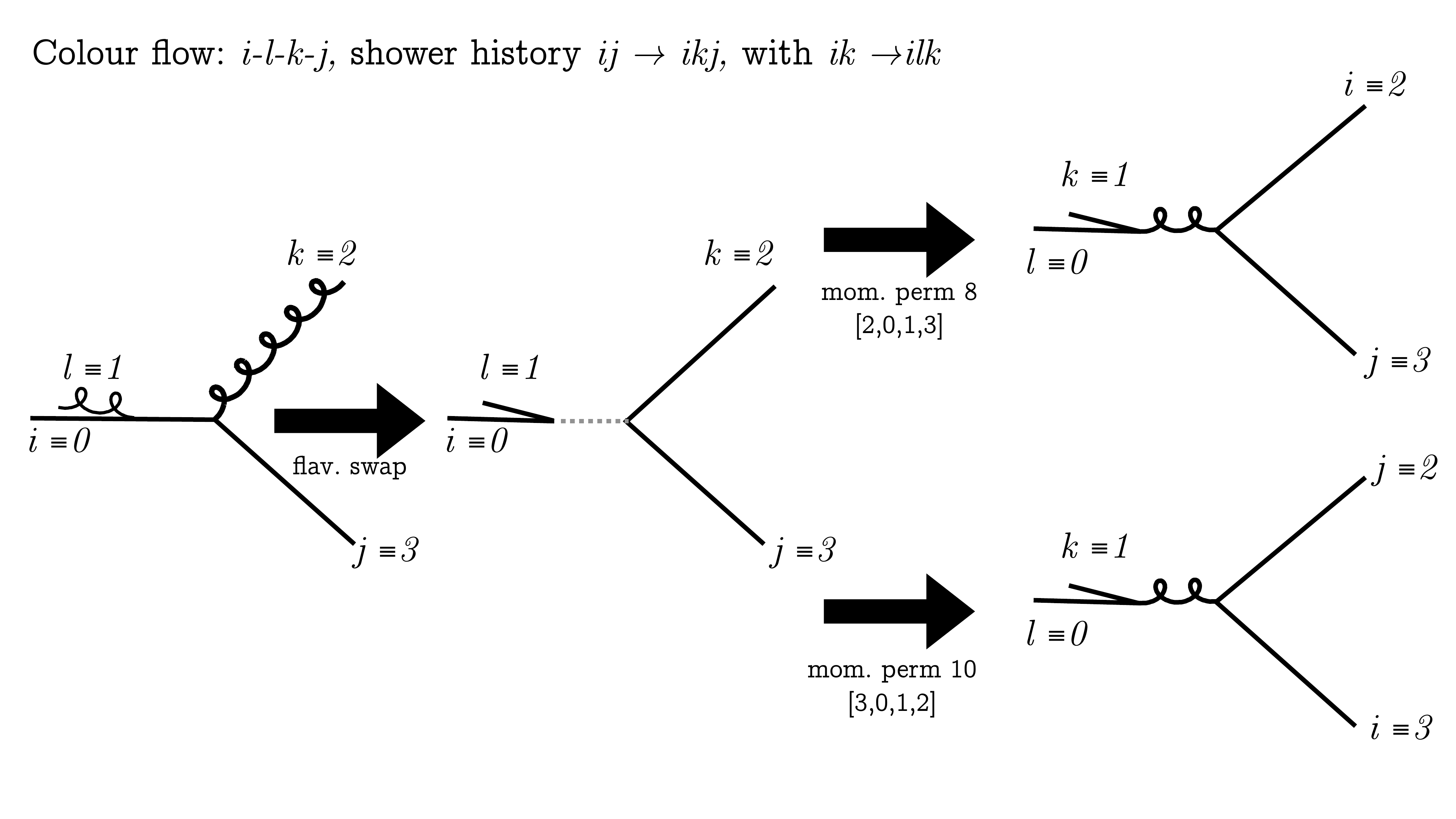}}
  {\footnotesize%
\begin{verbatim}
  shower_total = [2.888e+12, 1.667e+11]
  exact_total  = [2.858e+12, 2.168e+11]
  flav swap probability = 1.687e-02
  imom_perm =  0=[0,1,2,3] shower = [8.528e+11, 1.424e+01] exact = [8.528e+11, 2.501e+02]
  imom_perm =  1=[0,1,3,2] shower = [4.290e+11, 2.073e+00] exact = [4.290e+11, 8.038e+01]
  imom_perm =  2=[0,2,1,3] shower = [3.593e+03, 2.560e+02] exact = [2.270e+03, 2.501e+02]
  imom_perm =  3=[0,2,3,1] shower = [2.265e+02, 0.000e+00] exact = [1.519e+02, 9.291e-01]
  imom_perm =  4=[0,3,1,2] shower = [1.699e+03, 5.859e+01] exact = [1.150e+03, 8.038e+01]
  imom_perm =  5=[0,3,2,1] shower = [1.864e+02, 0.000e+00] exact = [1.279e+02, 9.291e-01]
  imom_perm =  6=[1,0,2,3] shower = [1.083e+11, 7.868e+00] exact = [1.083e+11, 5.997e+00]
  imom_perm =  7=[1,0,3,2] shower = [5.450e+10, 8.552e+00] exact = [5.450e+10, 2.044e+00]
  imom_perm =  8=[2,0,1,3] shower = [7.216e+11, 8.333e+10] exact = [7.066e+11, 1.084e+11]
  imom_perm =  9=[2,0,3,1] shower = [4.252e+02, 1.681e+01] exact = [1.785e+02, 2.044e+00]
  imom_perm = 10=[3,0,1,2] shower = [7.216e+11, 8.333e+10] exact = [7.066e+11, 1.084e+11]
  imom_perm = 11=[3,0,2,1] shower = [6.914e+02, 8.455e+00] exact = [3.014e+02, 5.997e+00]
\end{verbatim}
  }
  \caption{Illustration of
    the issue with PanLocal $2\times 12$ (flavour $\times$ momentum)
    swaps, for an $ilkj$ colour flow.
    The box illustrates the configuration of concern.
    The printout below the box illustrates shower and exact matrix
    element weights for the summed set of permutations and the individual
    momentum permutations.
    Pairs of numbers in square brackets are respectively for the $gg$
    and $q\bar q$ flavour assignments of the $lk$ pair.
    The permutation indicates which momentum index (running from $0$
    to $3$) is assigned to each of $i,l,k,j$.
    See text for further details.
  }
  \label{fig:pl-12-swap}
\end{figure}

We use Fig.~\ref{fig:pl-12-swap} as a guide.
We start off with the momentum arrangement on the left, with an $ilkj$
colour flow, with $il$ collinear, and the branching probability well
described by the shower.
In the computer printout that is below the diagrams, pairs of numbers
in square brackets represent the matrix element for the channels
where $l$ and $k$ are both gluons and that where they form a $q\bar q$
pair.
The configuration on the left in Fig.~\ref{fig:pl-12-swap}
corresponds to the $gg$ channel of momentum permutation 0.
In the singular limit being considered, the shower has the correct
weight for the $gg$ channel, which dominates hugely over the $q\bar q$
channel. 
That is illustrated in the printout, specifically that
for momentum permutation 0.
For each of ``shower'' and ``exact'', the two entries in each array
are for the middle two particles in the list, $lk$, to be a $gg$ or a
$q\bar q$ pair.
One sees that the shower has the correct $gg$ weight, and that the
$q\bar q$ weight is negligible in comparison.

The always-active swap algorithm starts by considering flavour, summed across
all allowed momentum permutations.
In some of the other momentum permutations, the $q\bar q$ flavour
assignment is significant, for example permutation $8$ (top right in
Fig.~\ref{fig:pl-12-swap}).
Furthermore, because we turn off spin correlations in the matching
step of the shower, the shower is missing an azimuthal modulation and
does not reproduce the correct matrix element.
This introduces a mismatch between shower and exact weights in the sum
across momentum permutations, which is visible in the
\ttt{shower_total} and \ttt{exact_total} lines of the printout.
The always-active swap algorithm therefore leads to swaps from the
$gg$ to the $q\bar q$ channel with non-negligible probability.
This is represented in the central configuration of the figure where
particles $l$ and $k$ have turned into a $q\bar q$ pair.
There is then a non-negligible probability to swap into other momentum
permutations, for example into permutation 8 (top right) or
permutation 10 (bottom right).

The combination of the always-active flavour swap and the momentum
swap resolves the issue of the missing spin correlations that would
have been obtained by direct branching into those right-hand
configurations.
However it introduces a different issue, related to the Sudakov form
factor.
Consider a 4-jet observable with $\beta=1$, like the $D$-parameter.
Its Sudakov form factor has double- and single-logarithmic parts.
The double-logarithmic part is universal across the different momentum
permutations.
However the single-logarithmic part changes according to the
permutation.
Swapping between permutations that have different single-logarithmic
behaviours mixes up those behaviours and so is expected to invalidate
the NLL accuracy of the shower.
Another perspective is that the ($2\times12$ always-active) swaps effectively
violate the PanScales conditions: a branching in a singular region
changes the overall hard event topology.

\begin{figure}
  \centering
  \includegraphics[width=0.9\textwidth]{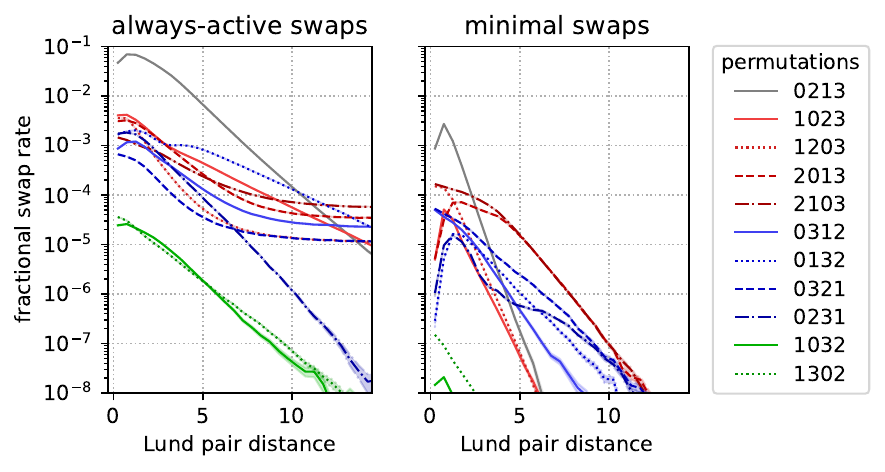}
  \caption{Fractional swap rate into different permutations, as a
    function of Lund-diagram distance between the two lowest
    transverse-momentum Lund declusterings.
    Shown for an ensemble of tree-level 4-parton events, fully matched
    to the leading-colour 4-parton matrix element.
    The left-hand plot shows the always-active swap scheme, with the
    plateau for some permutations indicative of the problem
    illustrated in Fig.~\ref{fig:pl-12-swap}.
    The right-hand plot shows the minimal swap scheme, demonstrating
    that all curves drop fast as a function of the Lund distance.
  }
  \label{fig:pl-swap-rates}
\end{figure}

To help convince ourselves that there is no similar issue hiding,
perhaps more subtly, in the minimal swap scheme, we carry out a
test where we generate tree-level 4-parton events with the PanLocal
phase-space map.
For each configuration we determine the distance in the Lund-diagram
between the two lowest transverse-momentum Lund declusterings (this distance
is invariant under swaps).
In Fig.~\ref{fig:pl-swap-rates}, as a function of that Lund distance, we
plot the fractional probability for swapping into a given permutation.
The left-hand plot is for the always-active scheme, and one sees that
some of the permutations have a swap probability that flattens out at
large Lund distance: this is a sign of swaps occurring with finite
probability no matter how singular the limit, for example into
permutation $2013$, as identified in the top right diagram of
Fig.~\ref{fig:pl-12-swap}.
The right-hand plot is for the minimal scheme.
There one sees that all swap probabilities vanish as the Lund
distance increases.
That is precisely the behaviour that we are looking for.
One also notes that the swap probabilities are in general much
smaller than in the always-active scheme.

\bibliographystyle{JHEP}
\bibliography{MC}

\end{document}